\documentclass[numberedappendix]{emulateapj}

\usepackage{graphicx}
\usepackage{bm}
\usepackage{epsf}
\usepackage{verbatim}
\usepackage{amsmath}
\usepackage{amsfonts}

\usepackage{epsfig}
\usepackage{dcolumn}
\usepackage{graphicx}
\usepackage{amssymb}

\usepackage{xcolor}
\definecolor{darkgreen}{rgb}{0.0,0.5,0.0}

\usepackage[colorlinks,linkcolor=darkgreen,citecolor=darkgreen]{hyperref} 

\usepackage[percent]{overpic}

\def\myfig#1{./#1}
\def\PlotFigs#1{#1} 

\def\PaperTwo{Gurwich \& Keshet 2017, in preparation}

\def\figsizeB{5.7cm} 

\newcommand{\ie}{\emph{i.e.,} }
\newcommand{\eg}{\emph{e.g.,} }
\newcommand{\cf}{\emph{cf.} }

\newcommand{\be}{\begin{equation}}
\newcommand{\ee}{\end{equation}}
\newcommand{\bea}{\begin{equation*}}
\newcommand{\eea}{\end{equation*}}
\newcommand{\beqr}{\begin{eqnarray} \nonumber}
\newcommand{\eeqr}{\end{eqnarray}}
\newcommand{\beqrb}{\begin{eqnarray}}
\newcommand{\eeqrb}{\nonumber \end{eqnarray}}
\newcommand{\fin}{\mbox{ .}}
\newcommand{\coma}{\mbox{ ,}}

\newcommand{\cm}{\mbox{ cm}}
\newcommand{\sr}{\mbox{ sr}}
\newcommand{\se}{\mbox{ s}}

\newcommand{\Myr}{\mbox{ Myr}}

\newcommand{\erg}{\mbox{ erg}}

\newcommand{\km}{\mbox{ km}}

\newcommand{\kpc}{\mbox{ kpc}}

\newcommand{\eV}{\mbox{ eV}}
\newcommand{\keV}{\mbox{ keV}}

\newcommand{\GeV}{\mbox{ GeV}}

\newcommand{\gama}{$\gamma$}

\newcommand{\grad}{\bm{\nabla}}
\newcommand{\vect}[1]{\bm{#1}}
\newcommand{\unit}[1]{\bm{\hat{#1}}}
\newcommand{\pr}{\partial}

\newcommand{\dgr}{{^\circ}}
\newcommand{\till}{{\mbox{--}}}

\newcommand{\myeta}{{\eta}}

\defcitealias{SuEtAl10}{S10}
\defcitealias{FermiBubbles14}{F14}
\newcommand{\Su}{{\citetalias{SuEtAl10}}}
\newcommand{\FT}{{\citetalias{FermiBubbles14}}}

\newcommand{\ApJMark}[1]{{#1}}

\begin{document}

\title{Fermi bubble edges: spectrum and diffusion function}

\shorttitle{Fermi bubble edges}

\author{Uri Keshet\altaffilmark{1} and Ilya Gurwich\altaffilmark{2}} 

\altaffiltext{1}{Physics Department, Ben-Gurion University of the Negev, POB 653, Be'er-Sheva 84105, Israel; ukeshet@bgu.ac.il}

\altaffiltext{2}{Department of Physics, NRCN, POB 9001, Beer-Sheva 84190, Israel}

\shortauthors{Keshet \& Gurwich}

\date{\today}

\begin{abstract}
Current measurements of the $\gamma$-ray Fermi bubbles (FB) are based on model-dependent tracers, carry substantial systematic uncertainties, 
and are at some tension with each other.
We show that gradient filters pick out the FB edges, which are found to smoothly connect to the bipolar X-ray structure emanating from the Galactic center, thus supporting the interpretation of the FBs as a Galactic-scale phenomenon.
The sharp edges facilitate a direct, model-free measurement of the peripheral FB spectrum.
The result is strikingly similar to the full FB-integrated spectrum, softened by a power law of index $\eta\simeq (0.2\mbox{--}0.3)$. This is naturally explained, in both hadronic and leptonic models, if cosmic rays are injected at the edge, and diffuse away preferentially at higher energies $E$.
The inferred, averaged diffusion function in the (more plausible) leptonic model, $D(E)\simeq 10^{29.5}(E/10\mbox{ GeV})^{0.48\pm0.02}\mbox{ cm}^2\mbox{ s}^{-1}$, is consistent with estimates for Kraichnan-like turbulence.
Our results, in particular the minute spatial variations in $\eta$, indicate that the FB edge is a strong, Mach $\gtrsim5$, forward shock.
\end{abstract}


\maketitle

\section{Introduction}
\label{sec:Intro}

Non-thermal lobes emanate from the nuclei of many galaxies.
These lobes, which are thought to arise from starburst activity or an outflow from a super-massive black hole \citep[for reviews, see][]{VeilleuxEtAl05,KingPounds15}, play an important role in the theory of galaxy formation \citep[\eg][and references therein]{Benson10}.
The presence of a massive, bipolar outflow in our own Galaxy has long been suspected, largely based on X-ray and radio signatures on large \citep{Sofue00}, intermediate \citep{BlandHawthornAndCohen03}, and small \citep{BaganoffEtAl03} scales, indicating an energetic, $\gtrsim 10^{55}\erg$ event \citep[][and references therein]{VeilleuxEtAl05}.

This scenario was revived by the discovery \citep[][henceforth \Su]{DoblerEtAl10, SuEtAl10} of two large \gama-ray, so-called Fermi, bubbles (FBs), symmetrically rising far from the Milky Way plane yet morphologically connected, at least approximately and in projection, to the intermediate-scale X-ray outflow features.
Due to their dynamical, nonthermal nature, and the vast energy implied by their presumed Galactic-scale distance, an accurate interpretation of the FBs is important for understanding the energy budget, structure, and history of our Galaxy.

The FBs, extending out to latitudes  $|b|\simeq 55^\circ$, show a flat, $\epsilon I_\epsilon \propto \epsilon^{-1}$ \ApJMark{photon} energy spectrum in the ($1$--$50$) GeV range, with luminosity  $\sim 4\times 10^{37}\erg\se^{-1}$ \ApJMark{if they indeed extend above the Galactic center} \citep[GC; \Su,][henceforth \FT]{Dobler12, HuangEtAl13, HooperSlatyer13, FermiBubbles14}.
They are also seen in the microwave band \citep{Dobler12, PlanckHaze13}, as the so-called microwave haze \citep{Finkbeiner04}, a residual diffuse signal surrounding the GC. This component has a somewhat harder spectrum, $\nu I_\nu\propto \nu^{-0.55\pm0.05}$, corresponding to a cosmic ray (CR) electron (CRE) spectrum $dN/dE\propto E^{-2.1\pm0.1}$ \citep[\eg][]{PlanckHaze13}.

The tentative identification of the FBs as massive structures emanating from the GC, rather than small lobes of a nearby object seen in (an unlikely; \eg \FT) projection, is based mainly on the coincidence of the lobe's base, to within a few degrees, with the GC.
The case would be rendered much stronger if the \ApJMark{FB edges could be firmly associated with the} bipolar X-ray structure on intermediate scales, clearly emanating from the GC; 
this is one of the goals of the present work.

Additional, less direct indications for the FB--GC association include the radio emission being too faint for a nearby source confined to the already-magnetized thick Galactic disk, the high emission measure of tentatively related high-latitude X-ray features \citep[][\ApJMark{but} see \ApJMark{discussion} below]{KataokaEtAl13}, and the low-latitude depolarization of linearly polarized lobes, partly but not \ApJMark{conclusively} 
coincident with the FBs \citep{CarrettiEtAl13}.
Stronger evidence that the FBs are a Galactic-scale phenomenon arises from the orientation of their symmetry axis being nearly exactly perpendicular to the Galactic plane: this is unlikely for the random pointing of a small body, but natural for an extended structure bursting out of the Galactic disk.

In spite of their dramatic appearance in the \gama-ray sky, the nature of the FBs is still debated.
Different models were proposed (\Su, \FT), interpreting
the FB edge as an outgoing shock \citep{FujitaEtAl13}, a termination shock of a wind \citep{Lacki14, MouEtal14}, or a discontinuity \citep{Crocker12, GuoMathews12, SarkarEtAl15};
the \gama-ray emission mechanism as either hadronic \citep{CrockerAharonian11, FujitaEtAl13} or leptonic \citep{YangEtAl13};
the underlying engine as a starburst \citep{CarrettiEtAl13, Lacki14, SarkarEtAl15}, an SMBH jet \citep{ChengEtAl11, GuoMathews12, ZubovasNayakshin12, MouEtal14}, or steady star-formation \citep{Crocker12};
and the CR acceleration mechanism as 
first-order Fermi acceleration, second-order Fermi acceleration \citep{MertschSarkar11, ChernyshovEtAl14}, or injection at the GC \citep{GuoMathews12, Thoudam13}.

More clues regarding the nature of the FBs have gradually surfaced.
Metal absorption lines of $\{-235,+250\}\km \se^{-1}$ velocities in the spectrum of quasar PDS 456, located near the base of the northern FB, \ApJMark{suggest} an outflow velocity $\gtrsim 900\km \se^{-1}$ \citep{FoxEtAl15}.
Longitudinal variations in the \ion{O}{7} and \ion{O}{8} emission line strengths, integrated over a wide latitude range \ApJMark{spanning} 
the full extent of both FBs, suggest a $\sim 0.4\keV$ FB multiphase plasma with a denser, slightly hotter edge, propagating through a $\sim 0.2\keV$ halo, thus suggesting a forward shock of Mach number $M=2.3_{-0.4}^{+1.1}$ \citep{MillerBregman16}.
By removing an FB template, \citet{SuFinkbeiner12} found a southeast--northwest, bipolar jet, with a cocoon on its southeast side; however, only the cocoon has so far been confirmed to be significant (\FT).

At the highest FB latitudes, an absorbed, $\sim0.3\keV$ X-ray component was reported \citep{KataokaEtAl13}, with a $\sim 60\%$ jump in the emission measure as one crosses outside the edge of the northern, but not the southern, bubble.
Such a jump would suggest that the FB edges are an $M\sim1.5$ reverse shock, terminating a wind.
However, the northern bubble is known to be much more prone to confusion and contamination than the southern bubble (\eg \FT), due to higher levels of dust and gas \citep[\eg][]{NarayananSlatyer16}, especially near the northeastern, Loop \ApJMark{I} feature where this X-ray jump was measured.
Indeed, the emission measure profile in the northeast sector shows substructure; in fact, a $20\%$ drop - rather than a jump - in emission measure can be seen at the putative edge of the FB \citep[figure 6 in][however, this edge is neither sharp nor well localized; see \S\ref{sec:Edges}]{KataokaEtAl13}.
More importantly, an $80\%$ emission measure drop can be seen across the more relaxed, southern edge.
We argue that this southern drop in emission measure (as one crosses outside the FB), combined with the much stronger drops at the supposedly related intermediate-latitude X-ray features \citep{BlandHawthornAndCohen03}, and the drop in the \ion{O}{8}/\ion{O}{7} line strength ratio \citep{MillerBregman16}, indicate that the FB edge is a forward, rather than a termination, shock.

The spectrum of the FBs, in \gama-rays and in \ApJMark{the} microwave, provides a major glimpse into the underlying physics.
Fully FB-integrated, so-called single-zone models, both hadronic and leptonic, have been put forward to explain the bubble-integrated spectra.
While reasonably fitting the data \citep[\FT;][and references therein]{NarayananSlatyer16}, these models have not been able to simultaneously account for both \gama-ray and microwave spectra without invoking ad hoc, physically unmotivated cutoffs on the CR spectrum.
In a separate publication (\PaperTwo), we argue that the only natural FB model fitting the integrated data is leptonic, and present a self-consistent, natural, single-zone model with no ad hoc cutoffs; a cooling break at $\sim 1\GeV$ photon energies then becomes evident.
Here we focus on the region immediately inside (\ie on the GC side of) the FB edges, where the spectrum can be robustly measured and easily interpreted.
Our present analysis applies to both hadronic and leptonic models.

There are some significant discrepancies between different previous measurements of the FB spectrum.
The integrated \ApJMark{flux} 
measured by {\FT} and \citet{HooperSlatyer13} is significantly higher than that in \cite{SuFinkbeiner12}.
Modest spectral variations in the \gama-ray spectrum were reported at high, $|b|>20\dgr$ latitudes \citep{HooperSlatyer13}.
In contrast, a significant hardening with increasing latitude, at least in the south bubble, was also reported \citep{YangEtAl14}, and modeled as the diffusion of CR ions (CRIs) injected near the GC in the framework of a hadronic FB model.
However, a later analysis (\FT) disagreed with both of these \citep{HooperSlatyer13, YangEtAl14} reports.
While the \gama-ray and microwave signals could generally be reconciled in a leptonic model (albeit with ad hoc cutoffs), the microwave signal was reported to harden in the mid-latitude range, $25\dgr\lesssim|b|\lesssim35\dgr$, \ApJMark{suggesting} 
some non-leptonic contribution \citep{NarayananSlatyer16}.

Such spectral measurements all relied on template reduction, where the observed sky in some wavelength band is modeled as the superposition of several tracers of the Galactic foreground, the extragalactic background, and an FB template.
These tracers include both observationally based maps, of gas, dust, Galactic synchrotron, and the cosmic microwave background (CMB), and synthetic distributions such as Galactic disk models, an idealized representation of Loop \ApJMark{I}, and simulations of Galactic CR propagation and radiation.
While this approach has been successful at separately identifying the FBs and the microwave haze, as evidenced by the morphological coincidence between the two, it can lead to spurious results, particularly in the spectrum, due to unknown components missing from the co-addition (such as arcs near Loop \ApJMark{I}, the cocoon, a putative mid-latitude hadronic component, and other substructure), correlations between the tracers, the choices of masks and templates (in particular, the precise definition of the FB template), etc.; see, for example \cite{KeshetEtAl04_EGRB}.
At least some of the above discrepancies were attributed to such caveats (\FT), resulting in substantial systematic uncertainties.

We adopt a different approach, based on the visible appearance of sharp edges in the \gama-ray map.
These edges are particularly evident in the south, where there is less substructure, in the east, where the projection appears to be favorable, and at higher energies, where foreground confusion diminishes. By comparing the \gama-ray brightness and its profile on each side of the edge, we robustly extract the spectrum, independent of any tracer or assumptions regarding the foreground and background.
Moreover, assuming that the FB edge is a shock, the spectrum we infer is closely related to the CR shock injection, with an evolution that can be easily modeled. The results are then used to critically test this shock assumption.

\begin{figure*}[t]
\PlotFigs{
\centerline{
\begin{overpic}[width=4.86cm]{\myfig{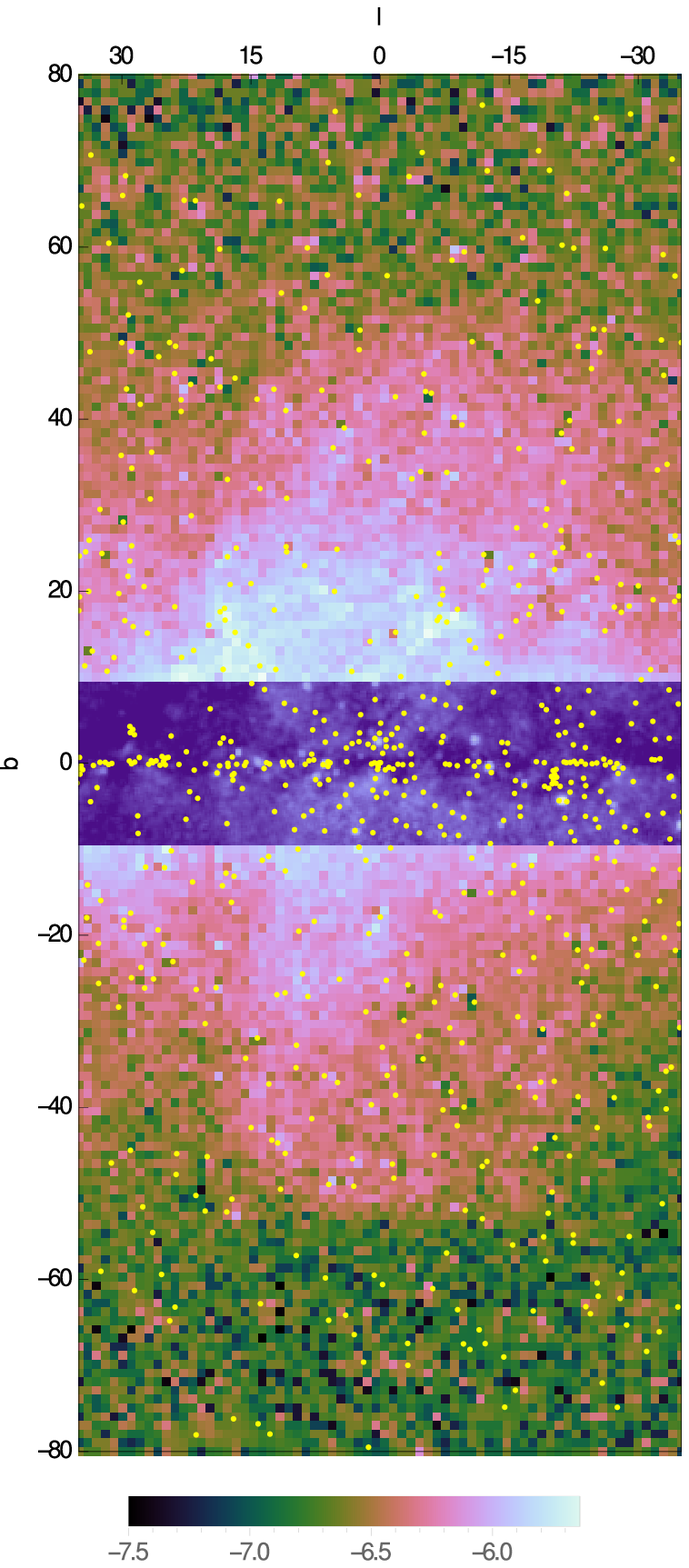}}
 \put (6,91) {\normalsize \textcolor{white}{(a)}}
\end{overpic}
\begin{overpic}[width=4.3cm, trim=0 -0.29cm 0 0]{\myfig{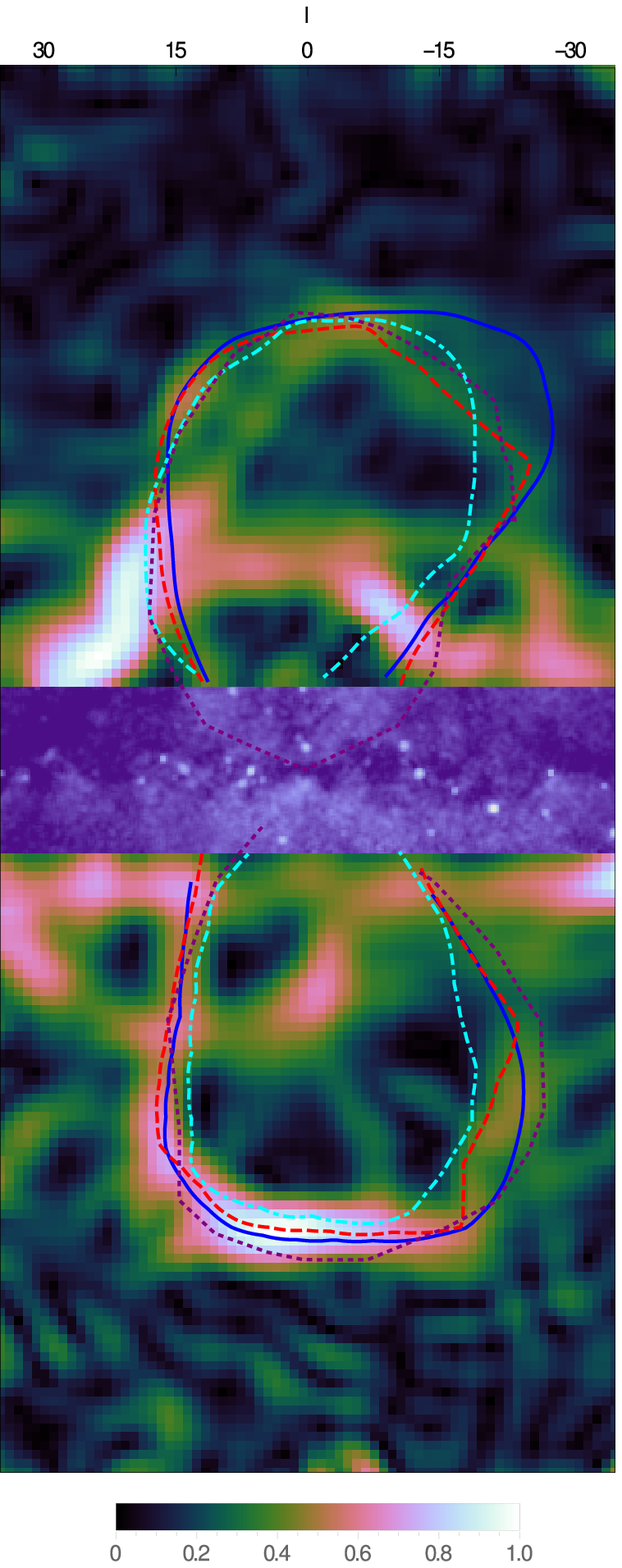}}
 \put (1,92) {\normalsize \textcolor{white}{(b)}}
\end{overpic}
\begin{overpic}[width=4.3cm, trim=0 -0.29cm 0 0]{\myfig{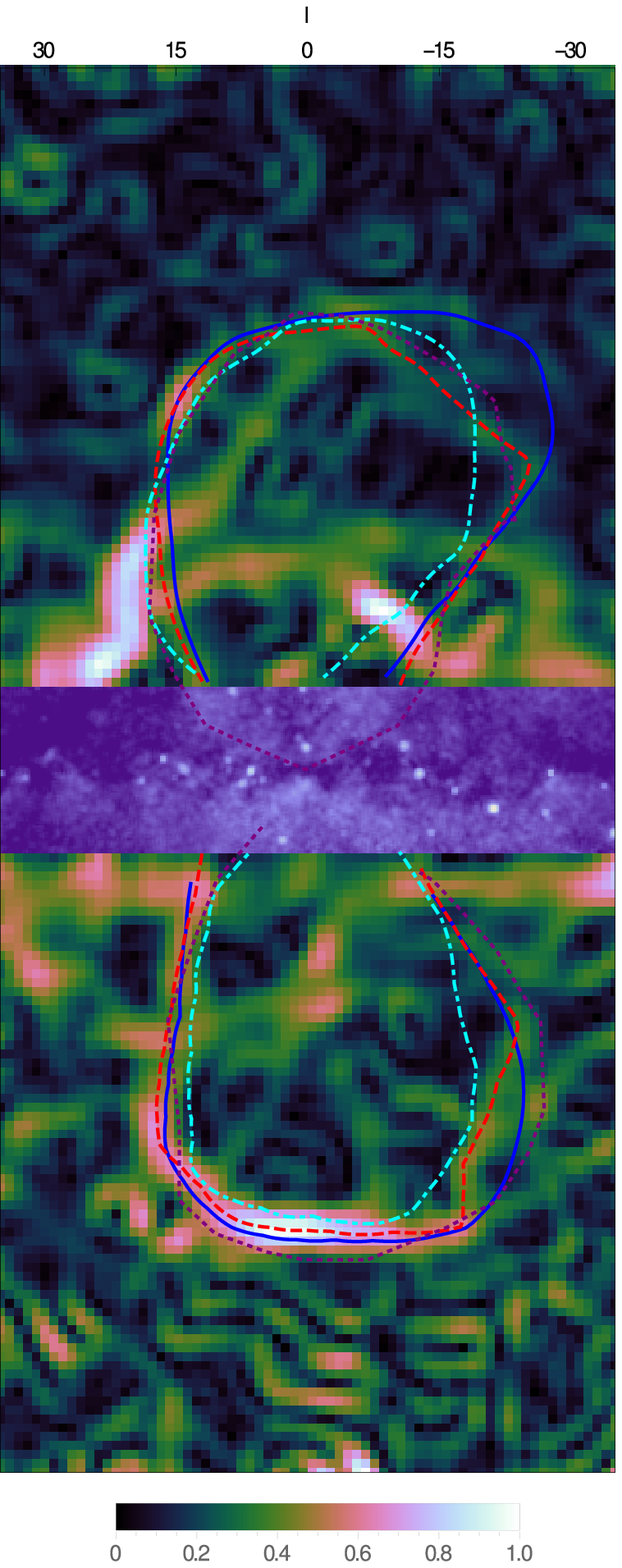}}
 \put (1,92) {\normalsize \textcolor{white}{(c)}}
\end{overpic}
\begin{overpic}[width=4.86cm]{\myfig{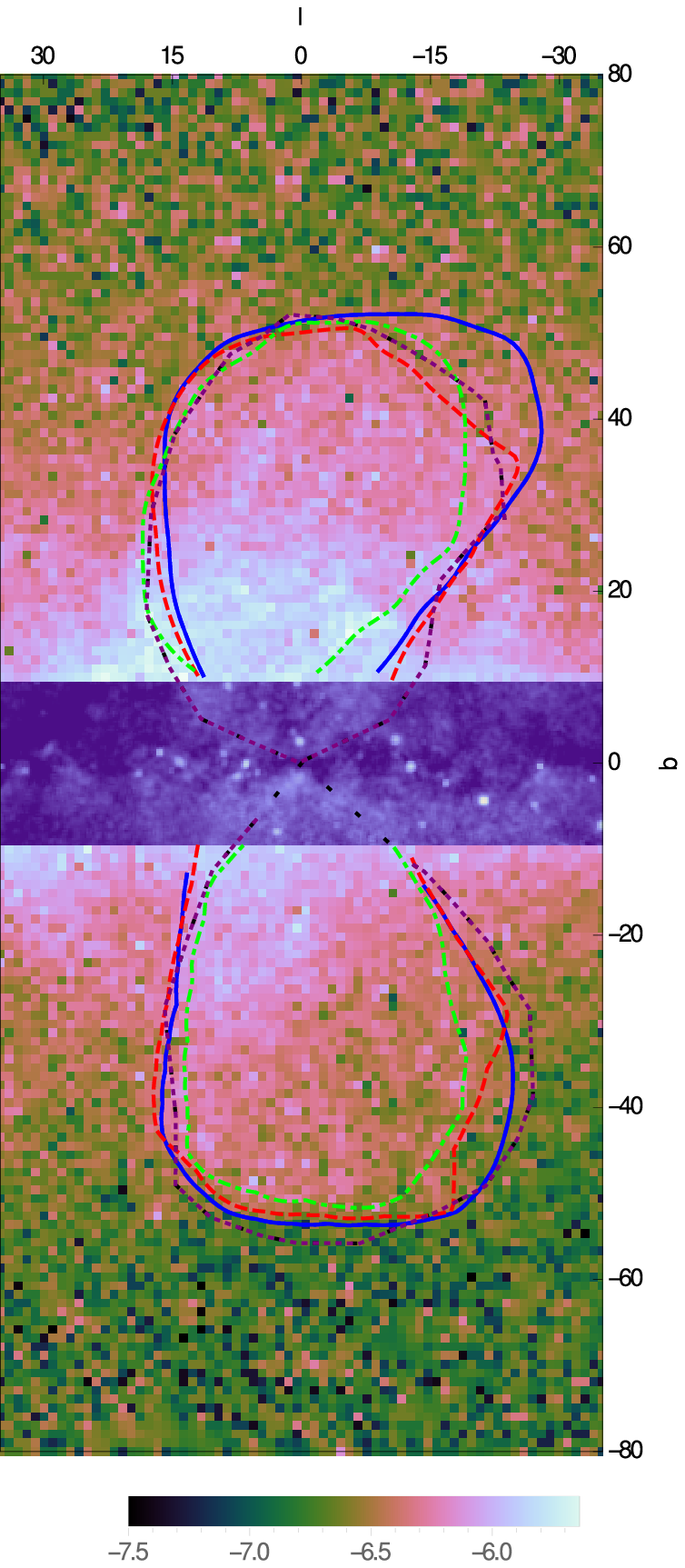}}
 \put (1,91) {\normalsize \textcolor{white}{(d)}}
\end{overpic}
}
}
\caption{
Illustration of the FB edge tracing method, shown in Galactic coordinates with a rectangular (\ApJMark{CAR}) projection and a cube-helix \citep{Green11_Cubehelix} color map, henceforth. The $|b|<10\dgr$ Galactic plane was masked and replaced by the \emph{ROSAT} \ApJMark{All Sky Survey band R6,} $(0.73-1.56)\keV$ map.
Panel (a): \emph{Fermi}-LAT (3-100) GeV all-sky map ($\log_{10}(F[\cm^{-2}\se^{-1}\sr^{-1}])$ color map). Point sources (yellow disks) were masked and compensated.
Gradient filters of $6\dgr$ (panel b) and $4\dgr$ (panel c) smoothing radii were applied to the Fermi map of panel (a), and are shown (on an arbitrary scale) with identified edge contours based on the $6\dgr$ filter (solid blue), the $4\dgr$ filter (dashed red), a manual trace (dot-dashed cyan), and {\Su} (dotted purple).
Panel (d): same as panel (a), with the four edge contours overlaid.
\label{fig:Edges}
}
\end{figure*}

The paper is arranged as follows.
In \S\ref{sec:Data}, we describe the preparation of \gama-ray data for this study.
The edges of the FBs are traced using gradient filters in \S\ref{sec:Edges}, and \ApJMark{subsequently} used to \ApJMark{radially bin the data in \S\ref{sec:Profiles} and} locally measure the FB spectrum in \S\ref{sec:Spectrum}.
The resulting spectrum is shown in \S\ref{sec:Tilt} to resemble a tilted, softer version of the full FB-integrated spectrum.
After deriving in \S\ref{sec:Model} a 'leaky box' model for the edge spectrum in the presence of CR diffusion, we estimate the diffusion function in \S\ref{sec:Diffusion}.
The results are summarized and discussed in \S\ref{sec:Discussion}.
Some convergence test and method variations are presented in Appendix \S\ref{sec:Convergence}.
We use $1\sigma$ error bars, unless otherwise stated, and Galactic coordinates, throughout.

\vspace{0.5cm}

\section{Data preparation}
\label{sec:Data}

We use the archival, $\sim 8$ year, Pass 8 Large Area Telescope (LAT) data from the Fermi Science Support Center (FSSC)\footnote{\texttt{http://fermi.gsfc.nasa.gov/ssc}}, and the Fermi Science Tools (version \texttt{v10r0p5}).
Pre-generated weekly all-sky files are used, spanning weeks $9\till422$ for a total of 414 weeks ($7.9$ year), with SOURCE class photon events.
A zenith angle cut of $90^\circ$ was applied to avoid CR-generated $\gamma$-rays originating from the Earth's atmospheric limb, according to the appropriate
FSSC Data Preparation recommendations.
Good time intervals were identified using the recommended selection expression \texttt{(DATA\_QUAL==1) and (LAT\_CONGIF==1)}.

Sky maps were discretized using an order 8 HEALPix scheme \citep{GorskiEtAl05}, providing an ample, sub-degree bin separation sufficient for the present analysis.
Event energies were logarithmically binned to cover the 100 MeV--1 TeV band.
Point-source contamination was minimized by masking pixels within $1^\circ$ of each point source in the LAT 4 year point-source catalog \citep[3FGL;][]{FermiPSC}, exceeding the $68\%$ containment of front-type events in all but the lowest (100--500) MeV band.
In order to reduce the Galactic foreground, we also masked the Galactic plane at $|b|<10^\circ$ latitudes.

\section{Edge tracing}
\label{sec:Edges}

We trace the FB edges using \emph{Fermi}-LAT events in the (3--100) GeV energy range, striking a good balance between bright FB emission, low Galactic foreground, and sufficient photon statistics.
The brightness map and edge tracing procedure are illustrated in Figure \ref{fig:Edges}.

To prevent masked point-source pixels from interfering with edge detection, we assign these masked pixels with the typical brightness of the surrounding, unmasked pixels.
As shown below, we are interested in edge transition widths spanning several degrees, much larger than the $1\dgr$ point-source masking radius.
Therefore, the details of this masking compensation have a negligible effect on our results.

Next, we smooth the map using a Gaussian filter of various angular scales $\psi$, and apply a standardized Bessel derivative kernel to pick up the strongest gradients on these scales. Extended edges that correspond to the FBs are clearly seen on scales $\psi\gtrsim2\dgr$, but become excessively smeared on scales $\gtrsim 8\dgr$.
Such a thickness range is consistent with previous studies (\Su, \FT), which traced the brightness profile along rays emanating from the estimated center of each bubble, and found most edge transitions to span ($2\dgr$--$6\dgr$) scales.

However, even in the $2\dgr<\psi<8\dgr$ range, the precise edge position varies as a function of $\psi$.
Figure \ref{fig:Edges} shows the edges extracted for $\psi=6\dgr$ (solid contours) and for $\psi=4\dgr$ (dashed contours). Also shown, for comparison, are the edges we pick by eye (dot-dashed contours) and the edge identified by {\Su} (dotted contour). The above four edges are respectively labeled 1 through 4, as detailed in Table \ref{tab:EdgeSummary}.

\begin{table}[h]
\begin{center}
\caption{\label{tab:EdgeSummary} Different FB edge contour tracing.
}
\begin{tabular}{cl}
\hline
Edge & Tracing Method \\
\hline
1 & Gradient filter on a $6\dgr$ scale \\
2 & Gradient filter on a $4\dgr$ scale \\
3 & Traced by eye \\
4 & Identified by {\Su} \\
\hline
\end{tabular}
\end{center}
\end{table}

As Figure \ref{fig:Edges} shows, the gradient filter-based edges 1 and 2 agree with each other to within $\sim2\dgr$, except at the northwest part of the northern bubble and the southwest part of the southern bubble, located symmetrically about the Galactic plane, where the two filters appear to pick up different physical features. Specifically, the \ApJMark{coarse}-grained edge 1 extends farther to the west at high latitudes; interestingly, the linearly polarized lobes \citep{CarrettiEtAl13} show a similar but much stronger tendency.
The manual edges 3 and 4 show stronger deviations, indicating that the potential for confusion with substructure and with sharp transitions is considerable.

To examine the connection between the FB edges and the intermediate-scale, bipolar X-ray features \citep{BlandHawthornAndCohen03}, in Figure \ref{fig:Edges} we replace the masked Galactic plane \ApJMark{at low latitudes} with the R6, $(0.73-1.56)\keV$ band of the \emph{ROSAT} all-sky X-ray background survey \citep{SnowdenEtAl97}, where these features are evident.
Edges 1 and 2, which are based on the gradient filters, appear to connect smoothly with these X-ray features, supporting the suspected common origin of these phenomena.

\section{\ApJMark{Radial profiles}}
\label{sec:Profiles}

\begin{figure}[h]
\PlotFigs{
\centerline{\epsfxsize=6cm \epsfbox{\myfig{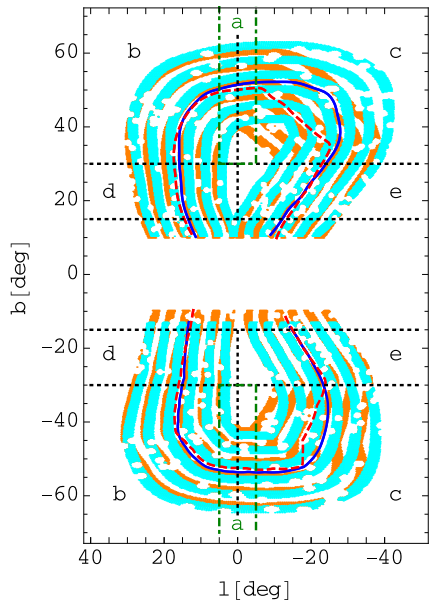}}}
}
\caption{ \label{fig:ThetaBins}
\ApJMark{
Illustration of the analysis sectors and the $\Delta\theta$ binning, in a rectangular projection. Sectors are labeled and delineated by lines (dot-dashed green for sectors a; dotted black for other sectors). Unmasked HEALPiX pixels are shown for $|\Delta\theta|$ intervals $(1,3)$, $(5,7)$, and $(9,11)$ degrees both inside and outside of edges 1 (the edge is a solid blue curve; pixels are bright cyan dots) and 2 (the edge is a dashed red curve; pixels are dark orange).
}
}
\end{figure}

\begin{figure*}[t]
\PlotFigs{
\centerline{
\begin{overpic}[width=5.7cm]{\myfig{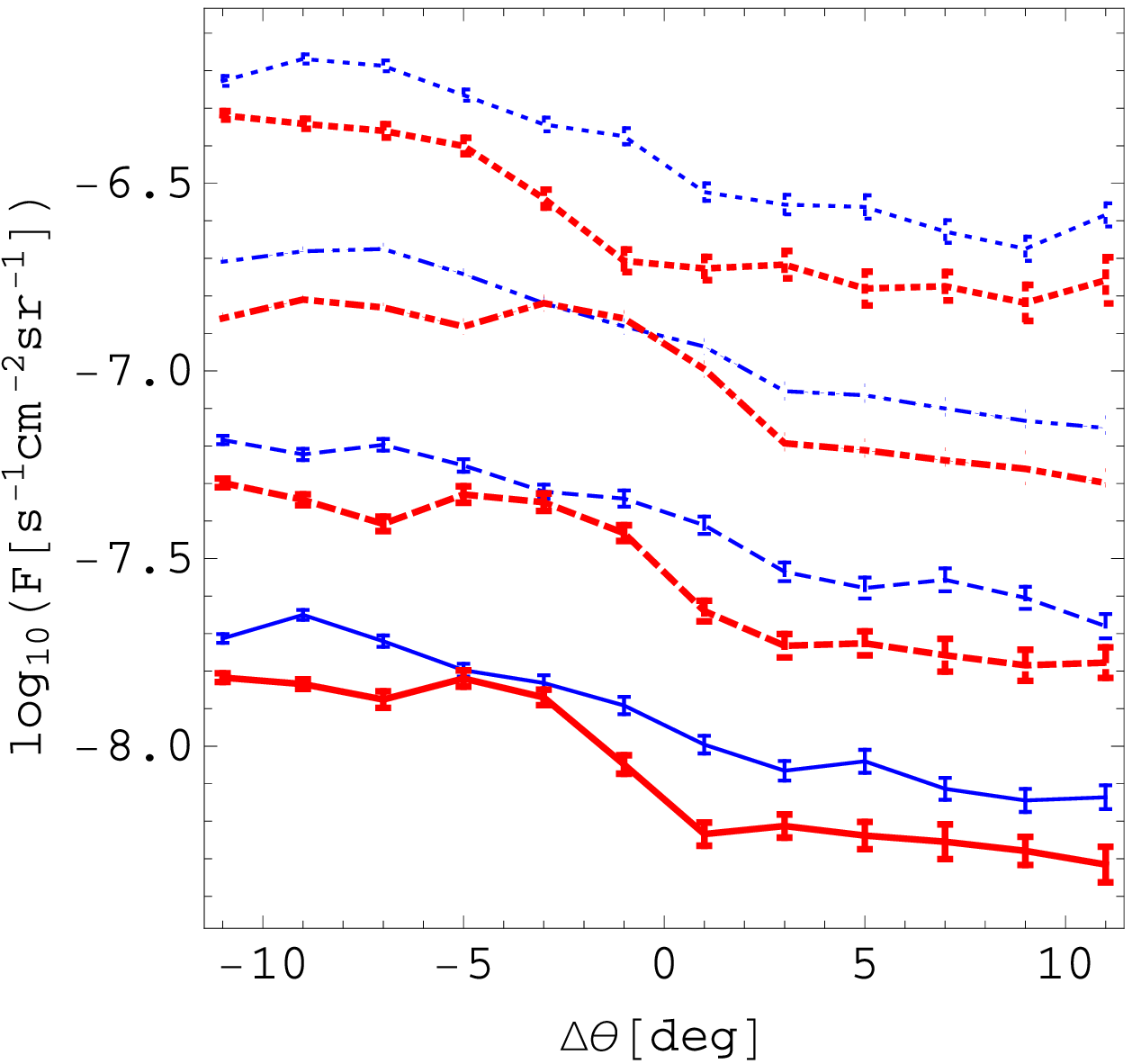}}
 \put (90,87) {\normalsize \textcolor{black}{a}}
\end{overpic}
}
\vspace{-4.0cm}
\centerline{
\begin{overpic}[width=5.7cm]{\myfig{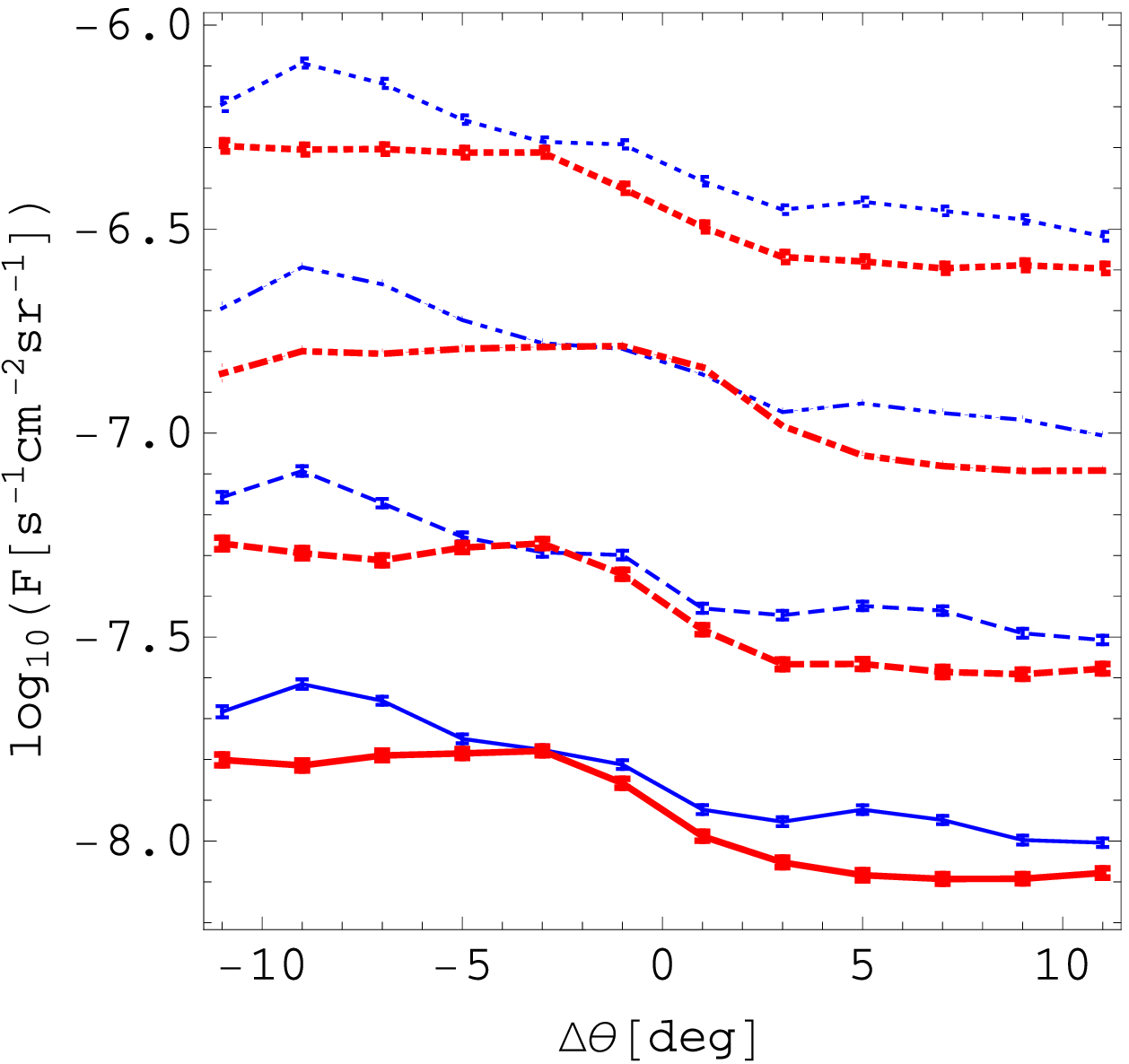}}
 \put (90,87) {\normalsize \textcolor{black}{b}}
\end{overpic}
\hspace{5.7cm}
\begin{overpic}[width=5.7cm]{\myfig{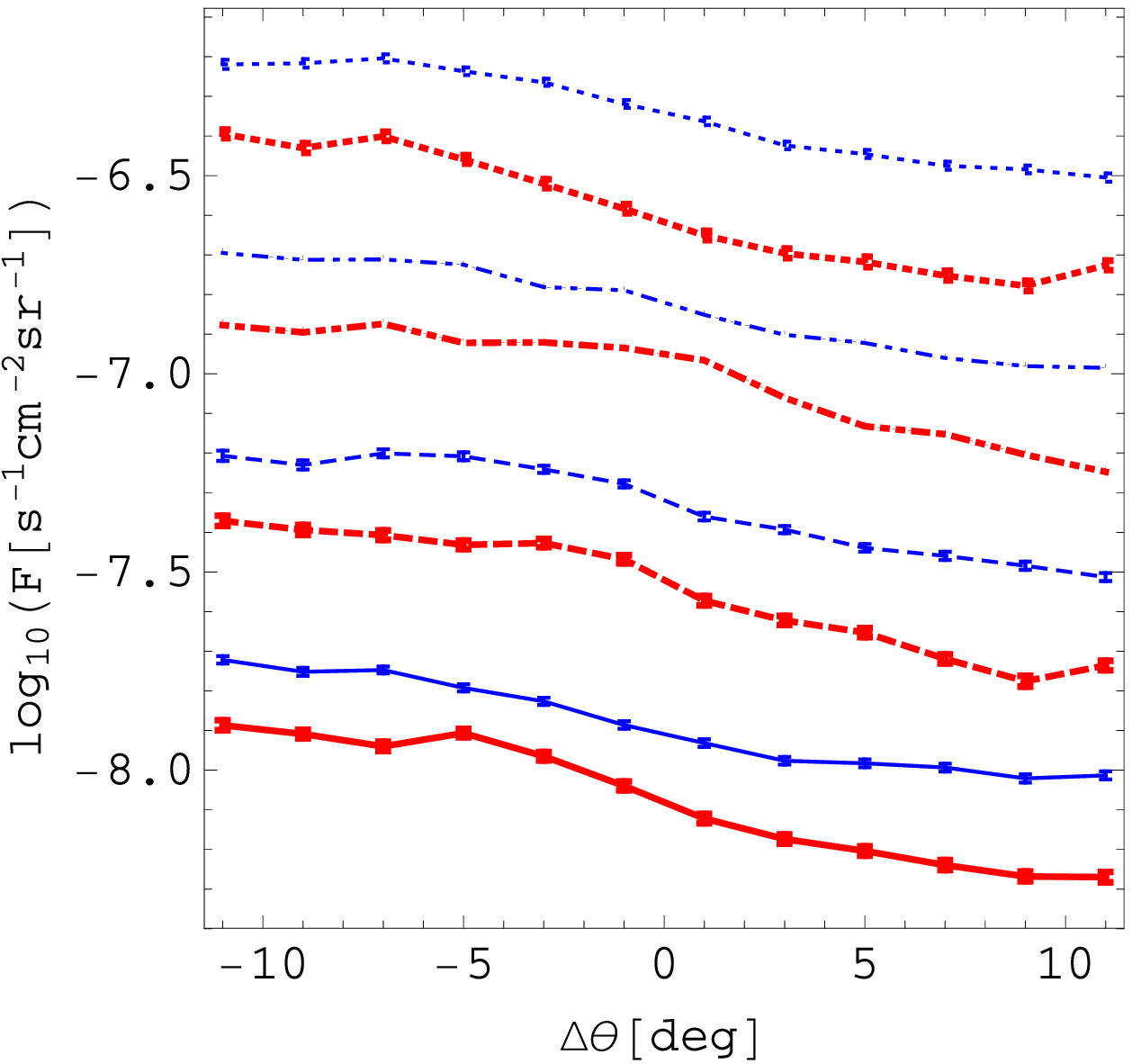}}
 \put (90,87) {\normalsize \textcolor{black}{c}}
\end{overpic}
}
\vspace{0.3cm}
\centerline{
\begin{overpic}[width=5.7cm]{\myfig{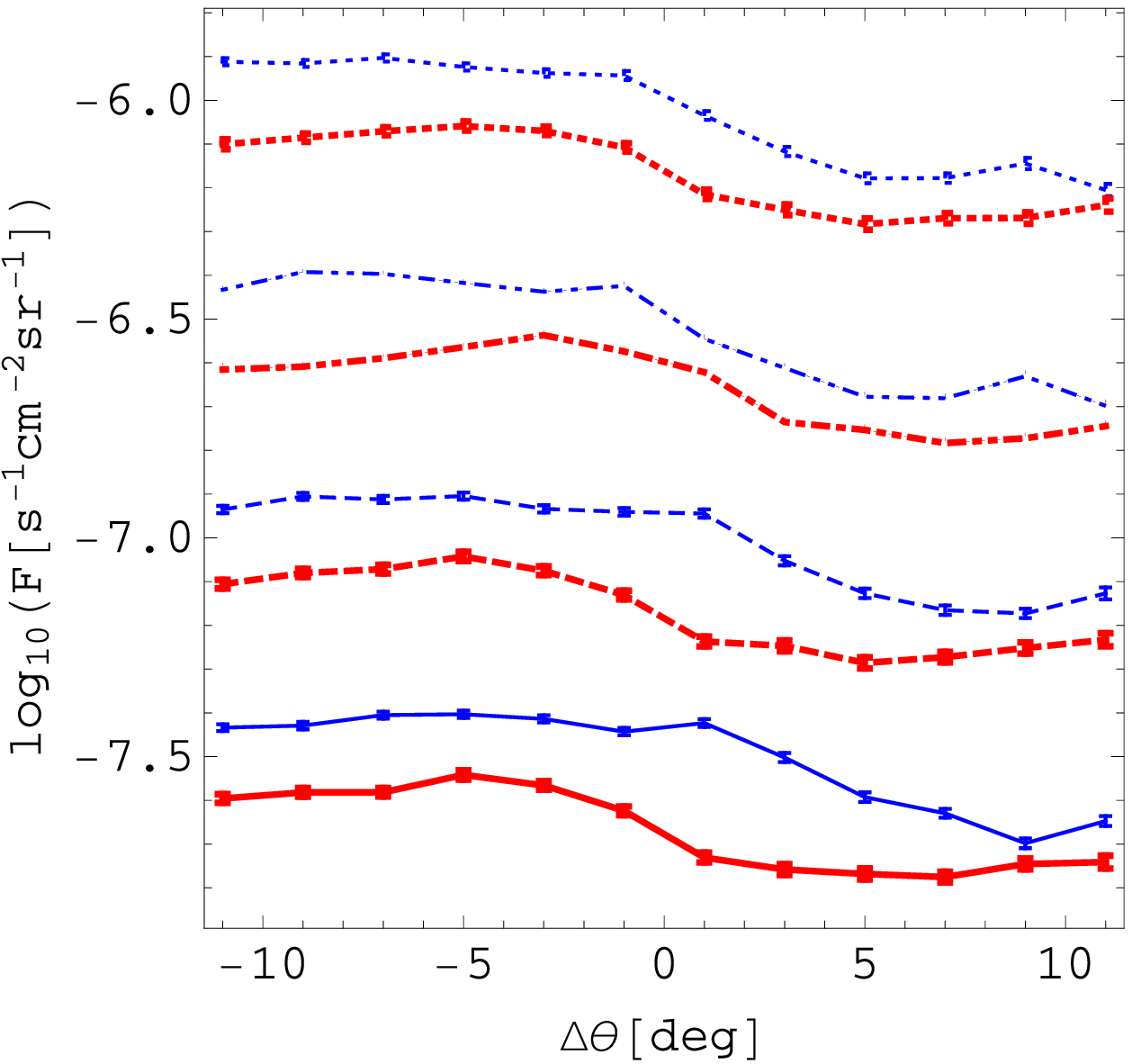}}
 \put (90,87) {\normalsize \textcolor{black}{d}}
\end{overpic}
\begin{overpic}[width=5.7cm]{\myfig{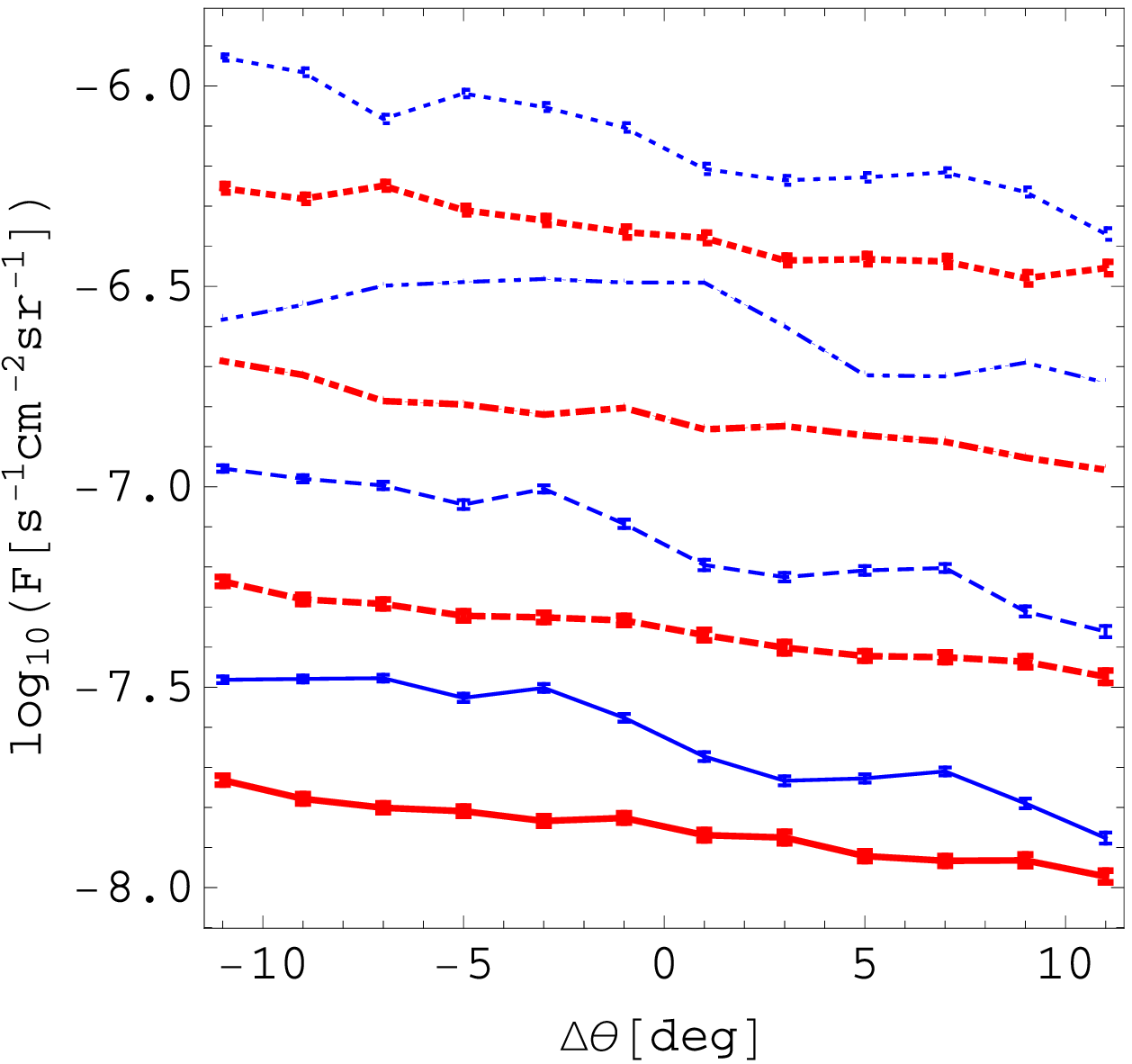}}
 \put (90,87) {\normalsize \textcolor{black}{e}}
\end{overpic}
}
%
%
}
\caption{
The north (blue symbols, with thin blue lines to guide the eye) and south (red, thick) radial profiles at (3--100) GeV energies, as a function of the angular distance from the edge, for the five sectors \ApJMark{in each hemisphere} (arranged roughly according to their locations along the northern bubble edge), defined by $|b|>30\dgr$ (top row) or $15\dgr<|b|<30\dgr$ (bottom row), and by $l>0\dgr$ (left column), $-5\dgr<l<5\dgr$ (middle panel in the top row) or $l<0\dgr$ (right column); each panel is labeled by its sector, as defined in Table \ref{tab:EdgeSectors}.
The four different edge choices are shown in each region (1--4; the line styles follow Figure \ref{fig:Edges}; edges 2--4 are shifted upward, for visibility, by increasingly large multiples of $10^{1/2}$).
\label{fig:RadialProfiles}
}
\end{figure*}

\ApJMark{To probe different parts of the FBs, we split each bubble into five sectors, as defined in Table \ref{tab:EdgeSectors}.
In each sector, and} for each traced edge, we compute the \gama-ray signal as a function of the shortest angular distance $\Delta \theta$ from the edge; positive (negative) $\Delta\theta$ values denote regions located outside (inside) the FB, \ie farther from (closer to) the GC.
\ApJMark{To reduce the noise level, we bin the HEALPiX pixels according to their $\Delta \theta$ values.
The sectors and the binning procedure are illustrated in Figure \ref{fig:ThetaBins}, depicting the definitions of the 10 sectors, and showing the unmasked pixels in several}
$\Delta\theta$
\ApJMark{intervals both inside and outside of both edges 1 and 2.}
The \ApJMark{resulting, binned radial profiles} are shown in Figure \ref{fig:RadialProfiles}, for each of the sectors.

\begin{table}[h]
\begin{center}
\caption{\label{tab:EdgeSectors} Different sectors along each bubble's edge.
}
\begin{tabular}{ccc}
\hline
Sector & Longitude Range & Latitude Range \\
\hline
a & $-5\dgr<l<5\dgr$ & $|b|>30\dgr$ \\
b & $l>0\dgr$ & $|b|>30\dgr$ \\
c & $l<0\dgr$ & $|b|>30\dgr$ \\
d & $l>0\dgr$ & $15\dgr<|b|<30\dgr$ \\
e & $l<0\dgr$ & $15\dgr<|b|<30\dgr$ \\
\hline
\end{tabular}
\end{center}
\end{table}

The edges found in the eastern sectors (b and d) and in the high-latitude sector (a) show a clear drop as one crosses outside the FB, with a typical $2\dgr$--$6\dgr$ transition width.
This drop tends to be clearer and better aligned with the edge in the southern bubble than it is in the northern hemisphere.
In most cases, a sharper or better aligned transition is found using edges 1 or 2, rather than with edges 3 or 4.
In the western sectors (c and e), the transition appears to be excessively smoothed or misaligned, except for low latitudes in the north bubble.

These results suggest that a good spectral measurement should be obtained using edges 1 or 2; for the north in sectors a, b, and e, and for the south in sectors a, b, and d. Possibly contaminated spectra may still be feasible, for the north in sector d, and for the south in sector c.
Sectors with excessively smoothed or slightly misaligned transitions may suffer from poor projection conditions. Here, extracting the spectrum from the edge region may underestimate the signal strength, but the spectral shape can still be captured if the signal is sufficiently strong and the foreground and background are sufficiently smooth.

\section{Spectral measurement}
\label{sec:Spectrum}

Next, we use the \ApJMark{four} edge contours \ApJMark{traced in \S\ref{sec:Edges}, as} defined in Table \ref{tab:EdgeSummary}, to measure the peripheral spectrum \ApJMark{of each of the FBs}, in \ApJMark{each of} the five sectors defined in Table \ref{tab:EdgeSectors}.

\begin{figure*}[t]
\PlotFigs{
\centerline{
\begin{overpic}[width=5.7cm]{\myfig{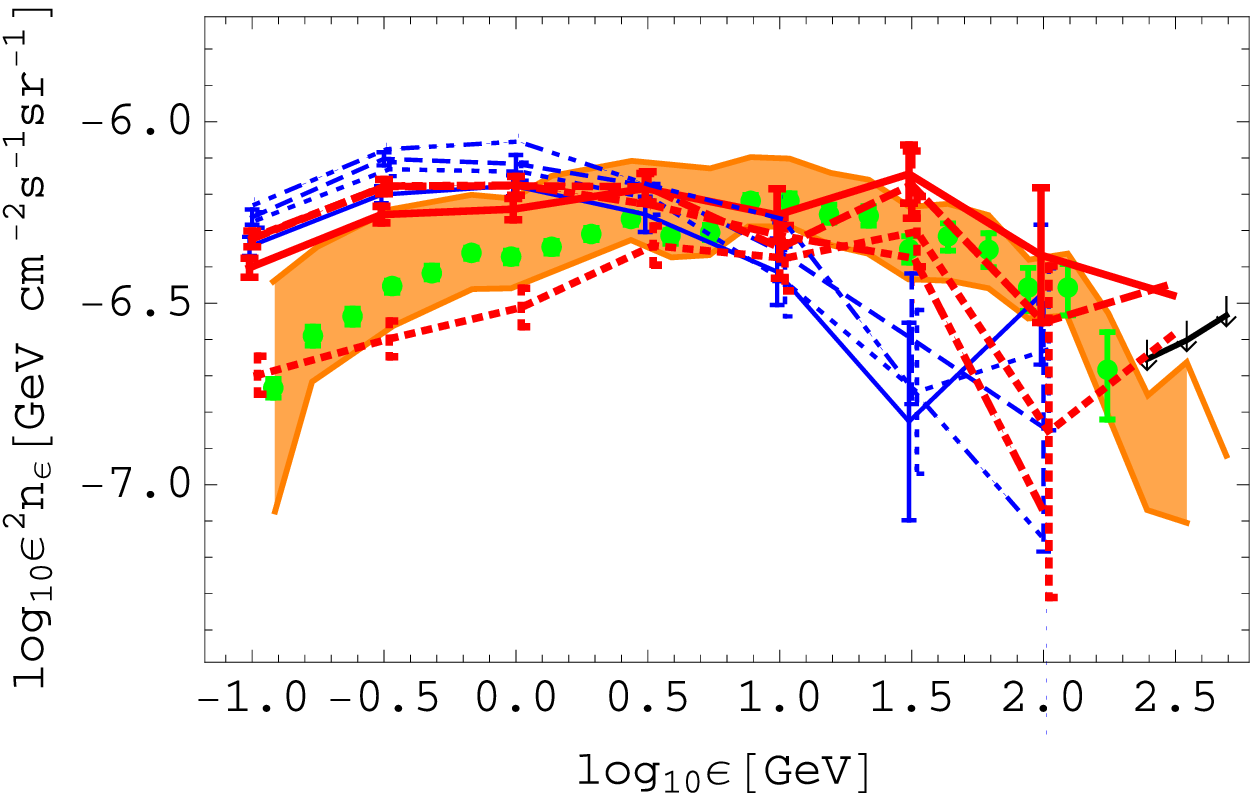}}
 \put (90,57) {\normalsize \textcolor{black}{a}}
\end{overpic}
}
\vspace{-2.7cm}
\centerline{
\begin{overpic}[width=5.7cm]{\myfig{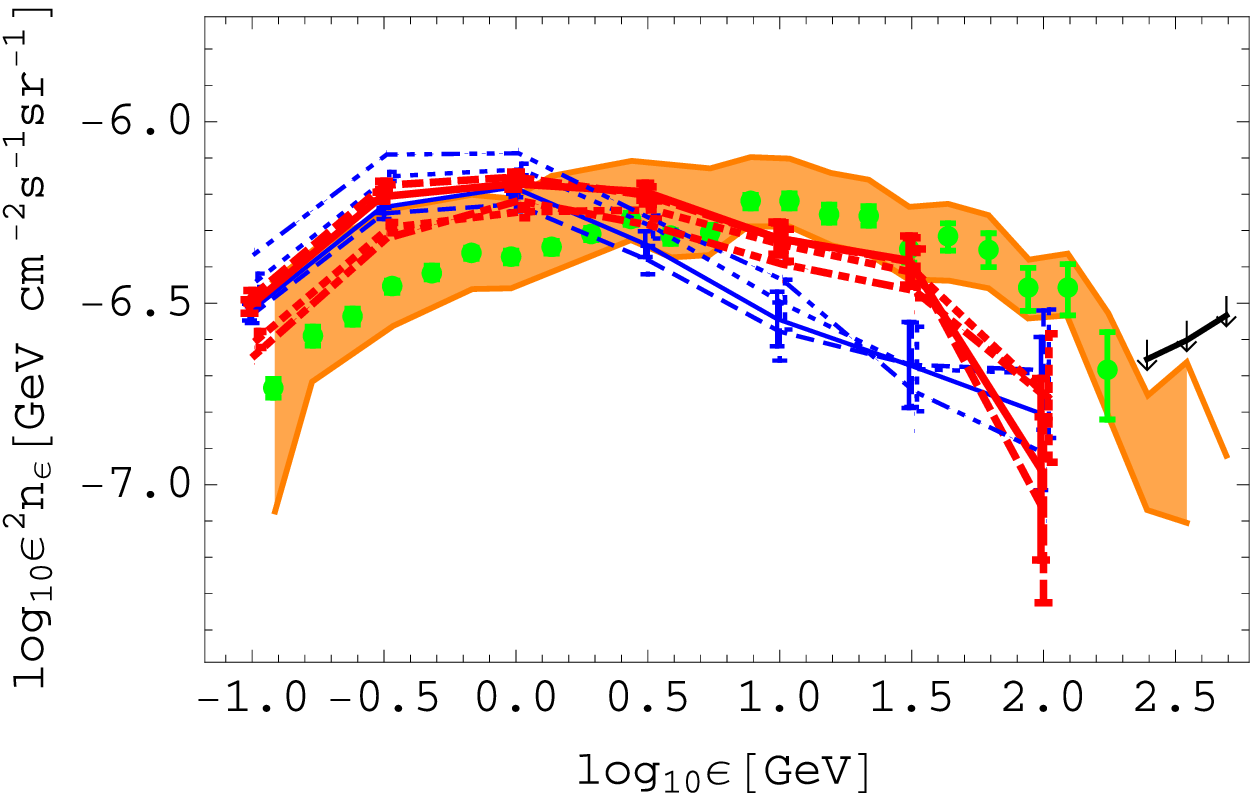}}
 \put (90,57) {\normalsize \textcolor{black}{b}}
\end{overpic}
\hspace{5.7cm}
\begin{overpic}[width=5.7cm]{\myfig{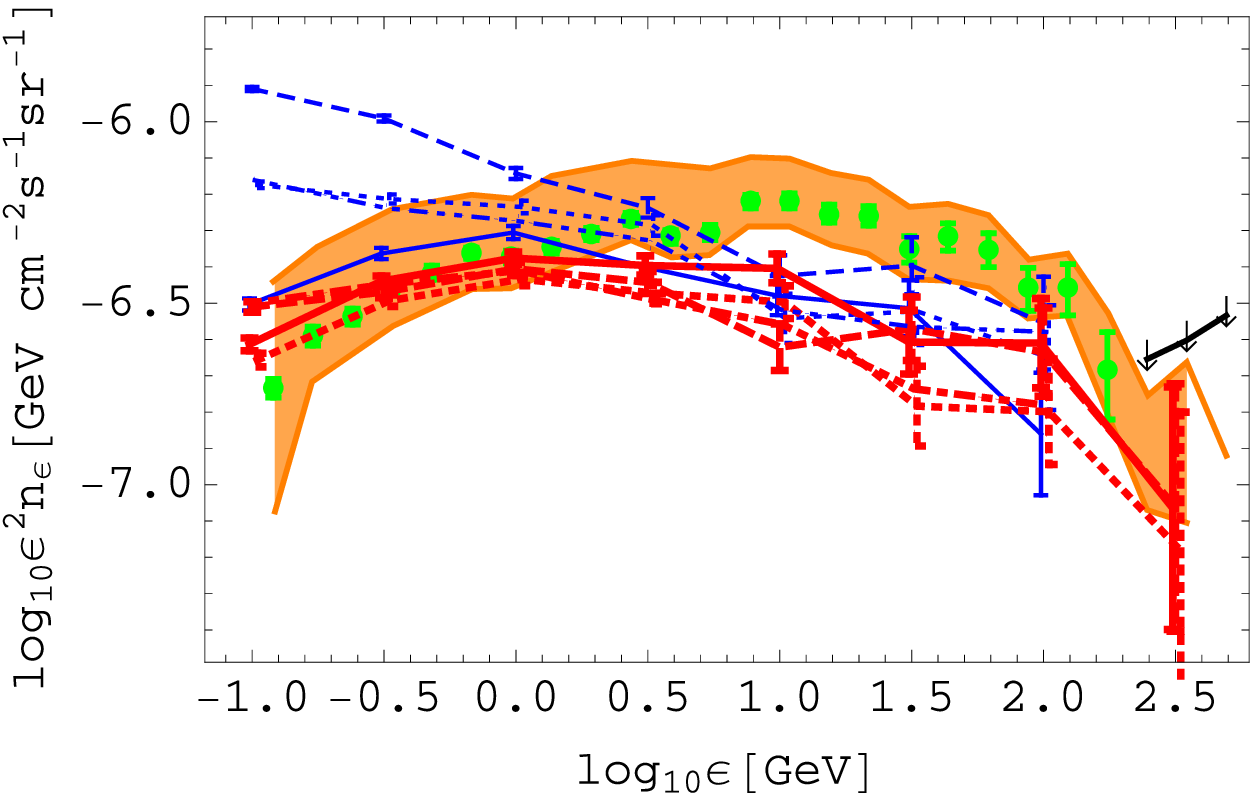}}
 \put (90,57) {\normalsize \textcolor{black}{c}}
\end{overpic}
}
\vspace{0.3cm}
\centerline{
\begin{overpic}[width=5.7cm]{\myfig{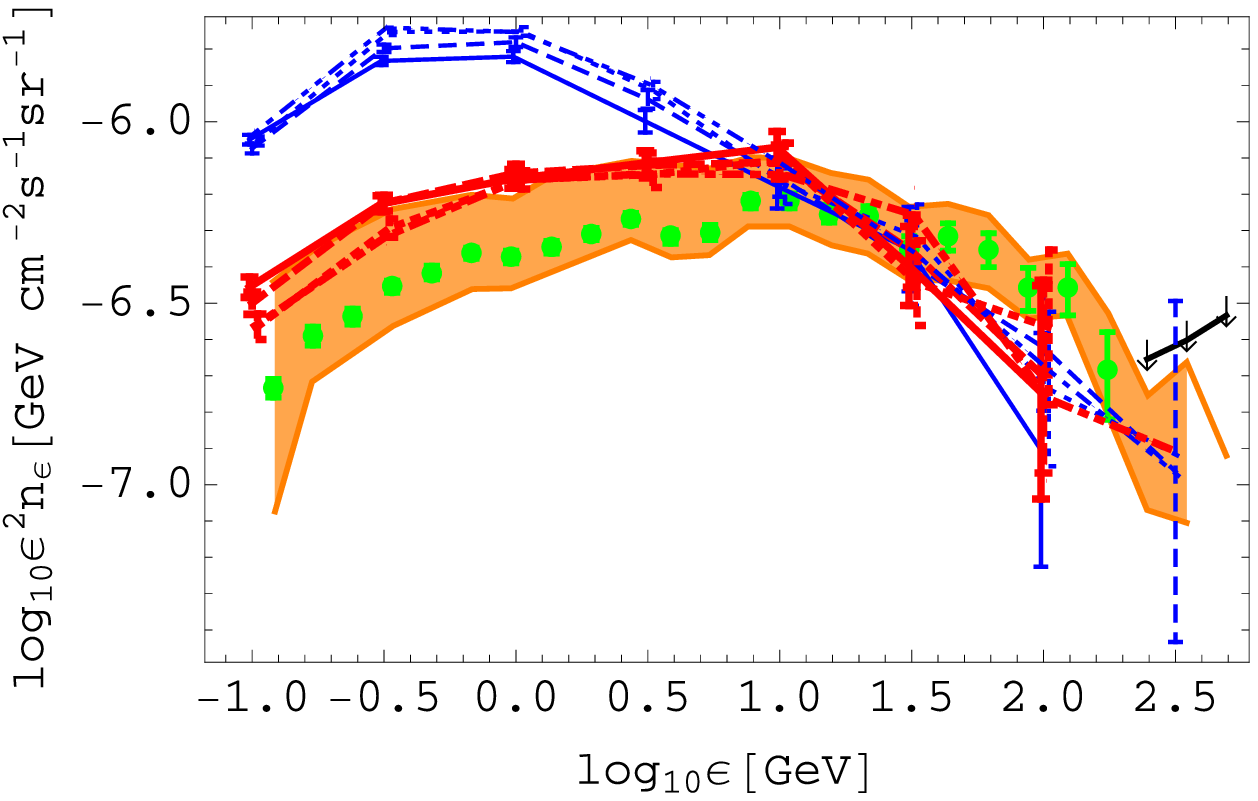}}
 \put (90,57) {\normalsize \textcolor{black}{d}}
\end{overpic}
\begin{overpic}[width=5.7cm]{\myfig{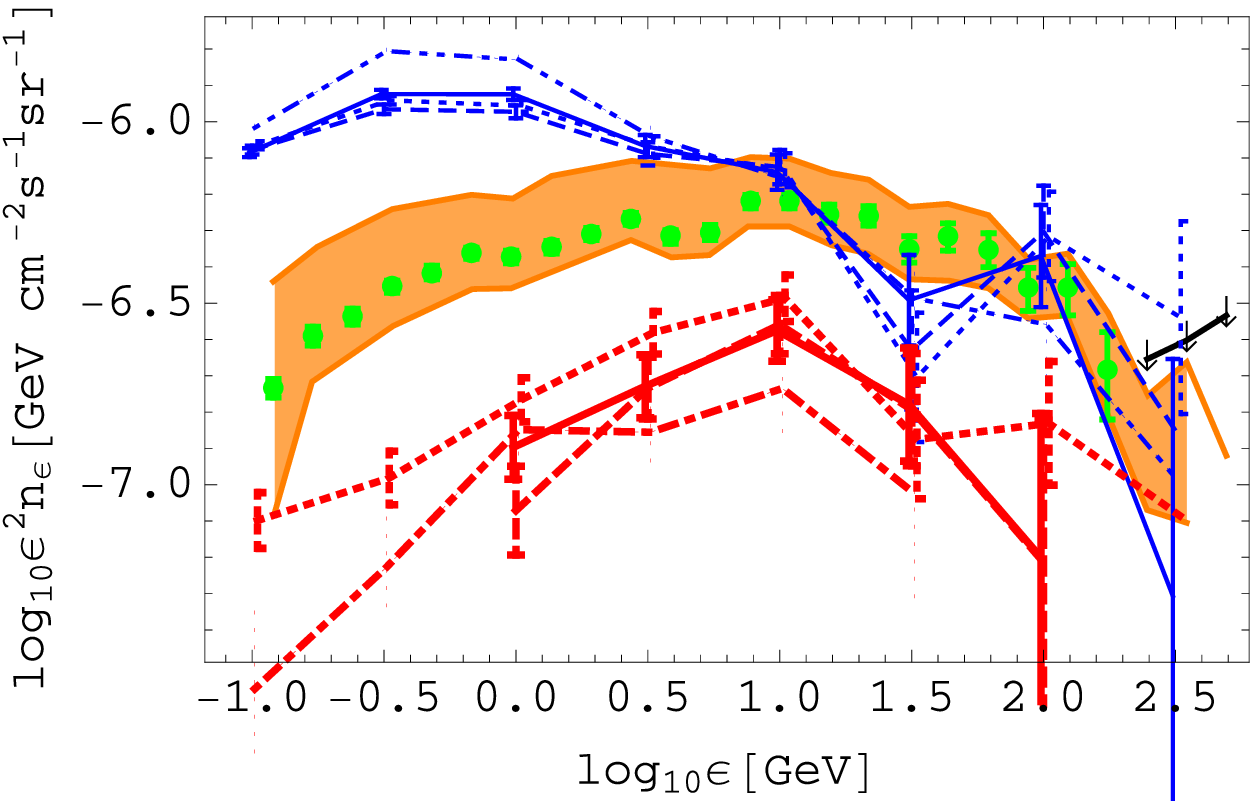}}
 \put (90,57) {\normalsize \textcolor{black}{e}}
\end{overpic}
}
}
\caption{\label{fig:DiffSpect}
Spectra of the five sectors \ApJMark{in each hemisphere}, 
obtained by subtraction across the edge, using a $\theta_g=6\dgr$ gap and a $\theta_r=3\dgr$ thick region on each side of the edge.
The symbols and curves are the same as in Figure \ref{fig:RadialProfiles}.
For comparison, we show the one-zone spectrum measured by {\FT} for a full-bubble template (statistical error bars are in green, systematic uncertainties are in shaded orange, and upper limits are black arrows).
}
\end{figure*}

In each \ApJMark{such} sector, we compute the signal in an outer region $S_o$, defined by $0<\Delta\theta-\theta_g/2<\theta_r$, and in an inner region $S_i$, defined by $-\theta_r<\Delta\theta+\theta_g/2<0$.
Here, $\theta_g$ is the angular gap between the inner and outer regions, corresponding to the $\sim (2\dgr$--$6\dgr)$ width of the transition.
The angular width of each region, $\theta_r\simeq (2\dgr$--$4\dgr)$, is chosen to be large enough to provide good photon statistics, yet small enough to minimize foreground contamination. Finally, the FB edge flux in each energy bin is estimated as $F_{\mbox{\scriptsize{edge}}}=F(S_i)-F(S_o)$.
\ApJMark{This provides an estimate of the FB contribution to region} $S_i$, \ApJMark{which is henceforth referred to as the edge region.}
The resulting spectra are illustrated in Figure \ref{fig:DiffSpect}, for $\theta_g=6\dgr$ and $\theta_r=3\dgr$; \ApJMark{the edge region here contains the second shaded bin inside the edge in Figure \ref{fig:ThetaBins}}.

The results confirm the expectation based on the radial profiles of Figure \ref{fig:RadialProfiles}.
Indeed, consistent spectra are obtained in most sectors, in particular when using edge 1.
In north sector c, all edges except edge 1 show evidence for contamination with a soft spectrum component.
We are unable to get a signal in south sector e, with any of the edges, possibly due to poor projection \ApJMark{conditions} that smooth the edge.
In the other sectors, the spectral shape is not highly sensitive to the choice of edge.
The spectra we find are also insensitive to the precise values of $\theta_g$ and $\theta_r$ (within reasonable ranges), and insensitive to foregrounds, as illustrated in \ApJMark{Appendix} \S\ref{sec:Convergence}.

Overall, the edge spectra are found to be qualitatively similar to the one-zone spectrum $F_{\mbox{\scriptsize{int}}}$ found by {\Su} and {\FT} for their full FB template.
This provides some model-independent support that the assumptions underlying the template removal methods are \ApJMark{not unreasonable}.
Naturally, the statistical errors are larger in our photon-limited method, especially at high energies.
Nevertheless, even accounting for the poor statistics of our $F_{\mbox{\scriptsize{edge}}}$ and the large systematic errors of the template-decomposition underlying $F_{\mbox{\scriptsize{int}}}$, 
\ApJMark{the data reveal a clear and significant difference between} $F_{\mbox{\scriptsize{edge}}}$ \ApJMark{and} $F_{\mbox{\scriptsize{int}}}$.

\begin{figure*}
\PlotFigs{
\centerline{
\begin{overpic}[width=5.7cm]{\myfig{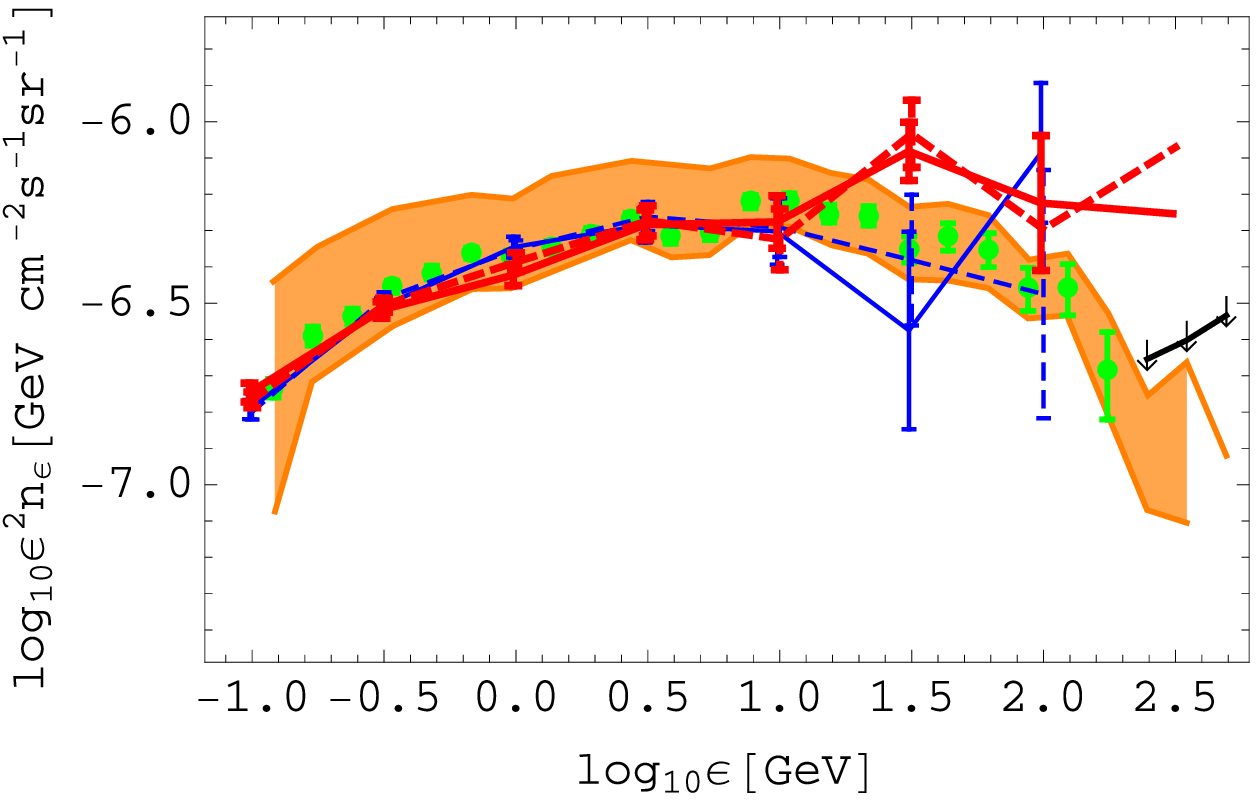}}
 \put (90,57) {\normalsize \textcolor{black}{a}}
\end{overpic}
}
\vspace{-2.7cm}
\centerline{
\begin{overpic}[width=5.7cm]{\myfig{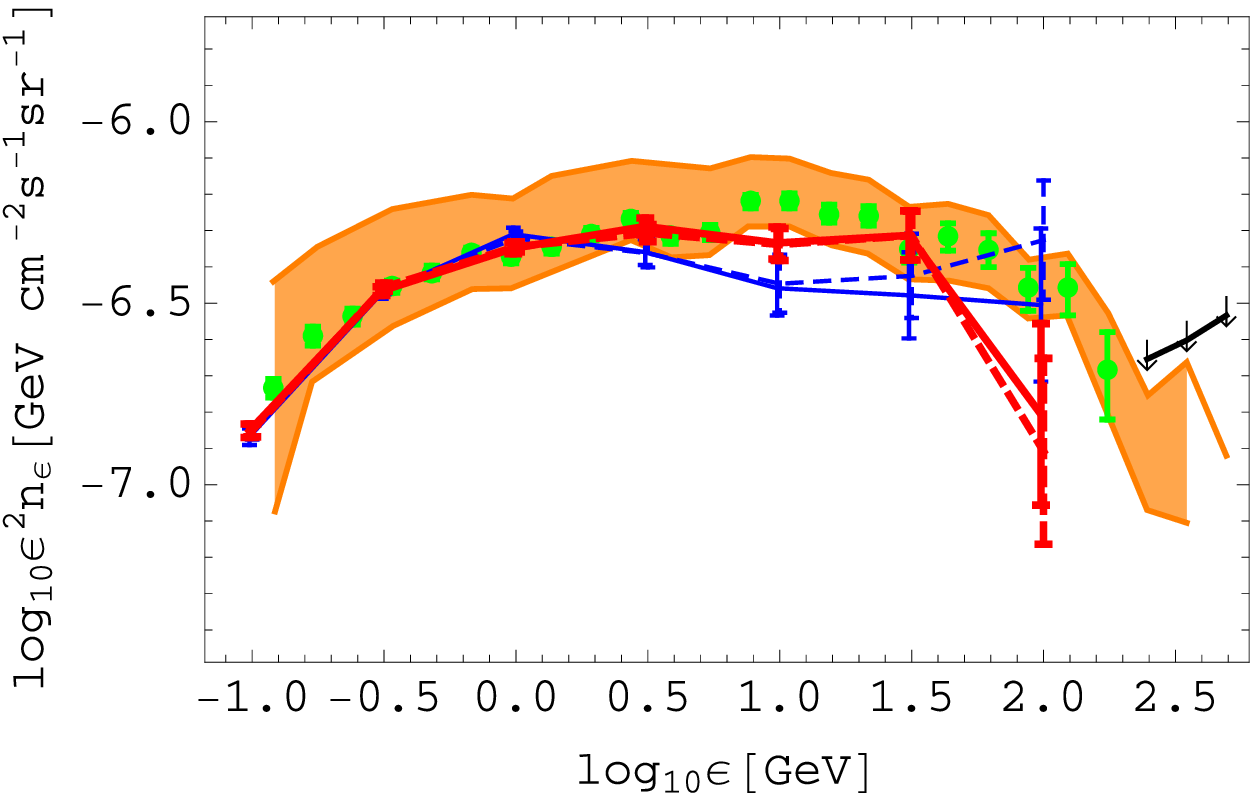}}
 \put (90,57) {\normalsize \textcolor{black}{b}}
\end{overpic}
\hspace{5.7cm}
\begin{overpic}[width=5.7cm]{\myfig{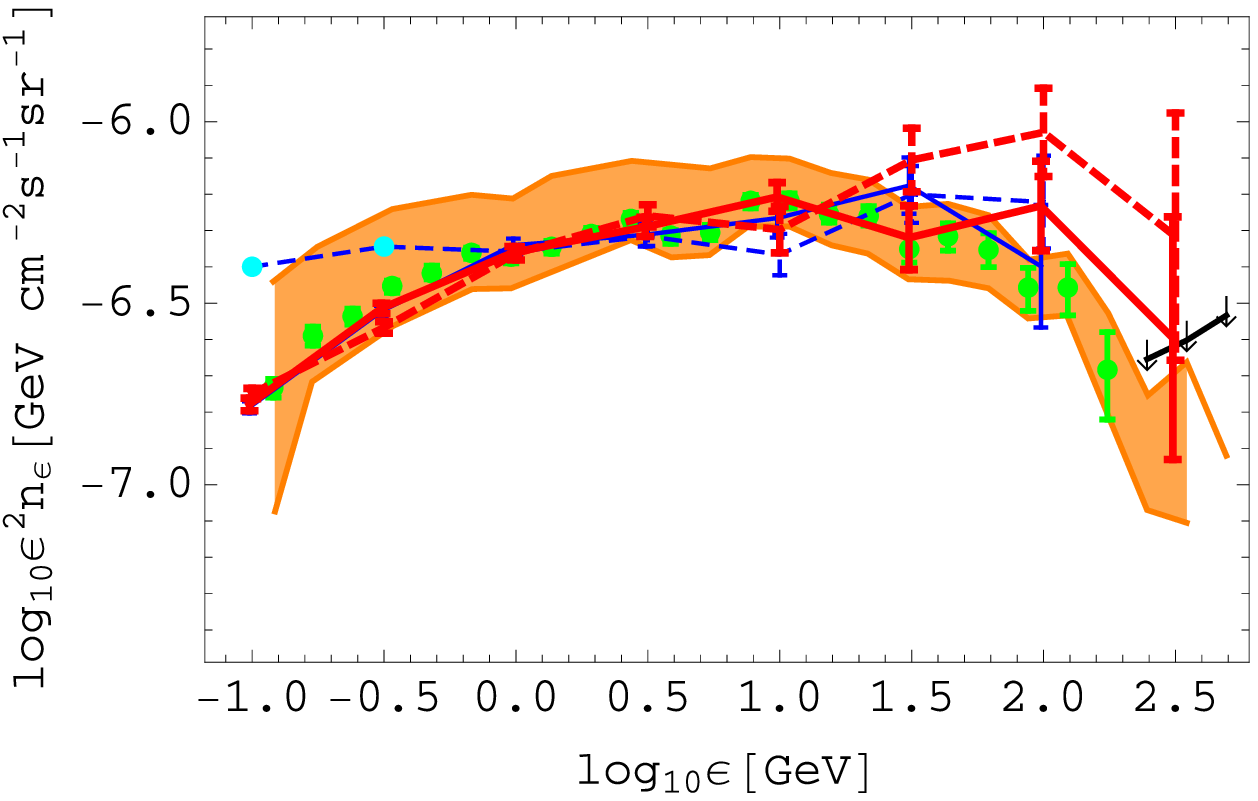}}
 \put (90,57) {\normalsize \textcolor{black}{c}}
\end{overpic}
}
\vspace{0.3cm}
\centerline{
\begin{overpic}[width=5.7cm]{\myfig{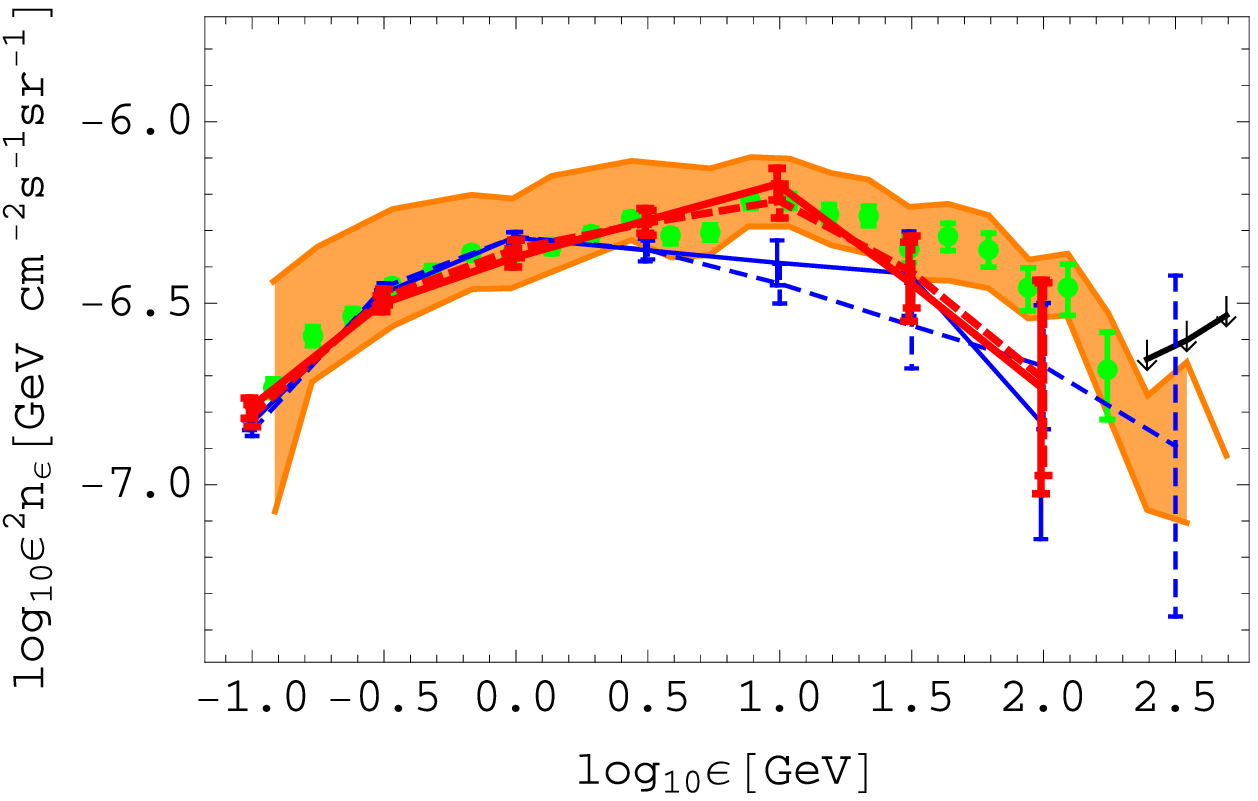}}
 \put (90,57) {\normalsize \textcolor{black}{d}}
\end{overpic}
\begin{overpic}[width=5.7cm]{\myfig{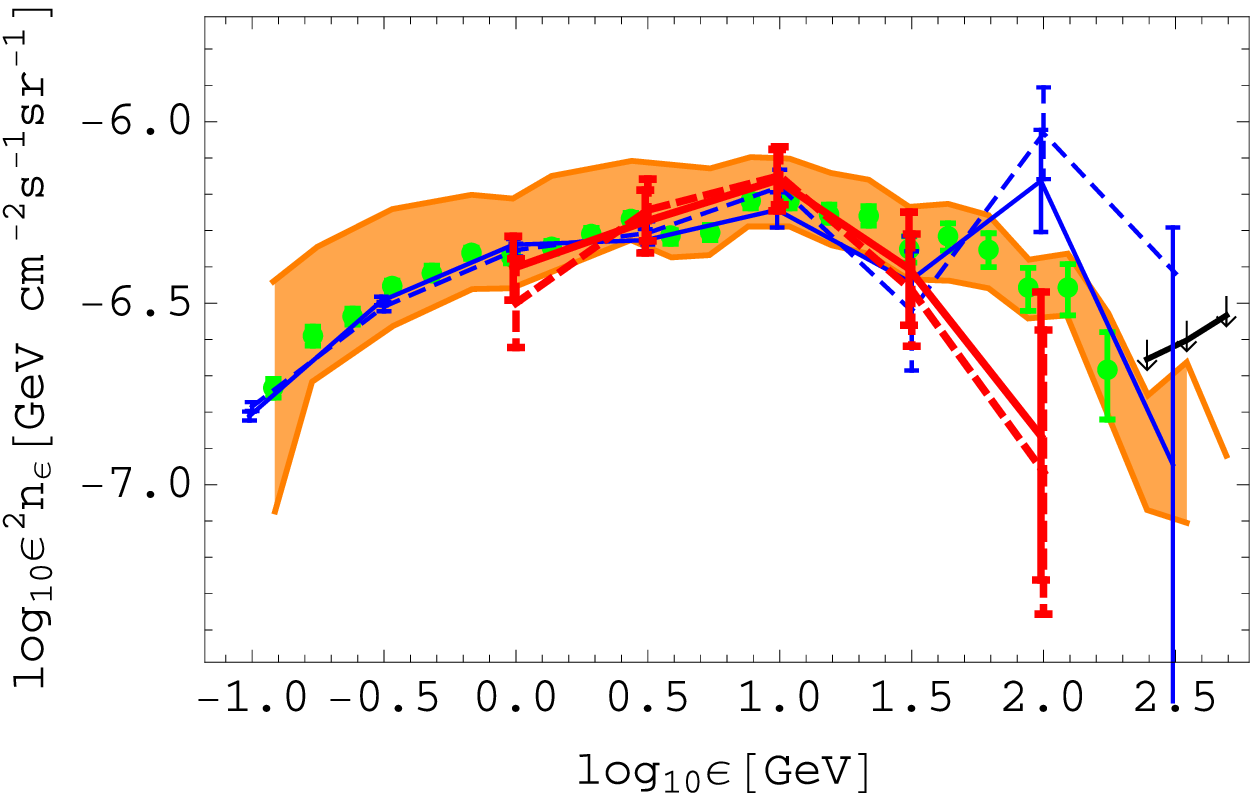}}
 \put (90,57) {\normalsize \textcolor{black}{e}}
\end{overpic}
}
}
\caption{ \label{fig:FitSpect}
Same as figure \ref{fig:DiffSpect}, but tilting the spectrum according to Eq.~(\ref{eq:Tilt}) with the parameters in Table \ref{tab:FitParams}.
For clarity, only two edges are shown in each sector: edge 1 (solid curves) and edge 2 (dashed).
}
\end{figure*}

\section{Spectral tilt}
\label{sec:Tilt}

Indeed, the spectra $F_{\mbox{\scriptsize{edge}}}$ we find near the FB edges \ApJMark{present} a significantly and consistently softer slope than that of the integrated spectrum $F_{\mbox{\scriptsize{int}}}$.
Moreover, we find that the difference in the logarithmic slope of the two is approximately energy-independent, nearly constant across the different sectors, and only slightly (albeit consistently) different among the two hemispheres.

\begin{figure}[h]
\PlotFigs{
\centerline{\epsfxsize=8cm \epsfbox{\myfig{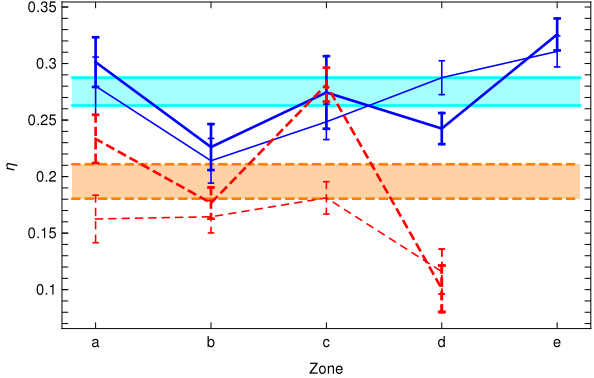}}}
}
\caption{ \label{fig:DiffusionFit}
The fit value of the tilt slope $\myeta$ in the different sectors for the north (blue error bars with solid blue lines to guide the eye) and south (red, dashed), using edge 1 (thick) and edge 2 (thin).  The mean over all sectors in each hemisphere is shown (the $95\%$ confidence intervals are displayed as shaded regions),
assuming it is constant in each bubble.
}
\end{figure}

To show this, we fit our edge spectrum to the integrated {\FT} spectrum by tilting the former, \ie multiplying it by a fixed power-law correction.
This may equivalently be written as a tilt applied to our edge spectrum,
\begin{equation} \label{eq:Tilt}
F_{\mbox{\scriptsize{int}}} \simeq F_{\mbox{\scriptsize{tilt}}} \equiv (\epsilon/1\GeV)^{\myeta} A F_{\mbox{\scriptsize{edge}}} \coma
\end{equation}
where $\epsilon$ is the photon energy, and $\{A,\myeta\}$ are fit parameters taken as constants, at least in each sector.
The resulting fit is shown for all five sectors \ApJMark{and both FBs} in Figure \ref{fig:FitSpect}.
The fit parameters are provided in Table \ref{tab:FitParams}, and the $\myeta$ values are additionally plotted in Figure \ref{fig:DiffusionFit}.
Due to the soft-spectral contamination of edge 2 in north sector c, we exclude the two lowest-energy data points (colored cyan in Figure \ref{fig:FitSpect}) in its specific fit.

\begin{table}[h]
\begin{center}
\caption{\label{tab:FitParams} Spectral tilt parameters for edges 1 and 2.
}
\begin{tabular}{ccccc}
\hline
Edge/ & North & North & South & South \\
Sector & $\log_{10}A$ & $\myeta$  & $\log_{10}A$ & $\myeta$ \\
\hline
1a & $-0.17\pm0.02$ & $0.28\pm0.03$ & $-0.18\pm0.01$ & $0.16\pm0.02$ \\
1b & $-0.13\pm0.01$ & $0.21\pm0.02$ & $-0.18\pm0.01$ & $0.16\pm0.01$ \\
1c & $-0.03\pm0.01$ & $0.25\pm0.02$ & $0.02\pm0.01$ & $0.18\pm0.01$ \\
1d & $-0.50\pm0.01$ & $0.29\pm0.01$ & $-0.22\pm0.01$ & $0.12\pm0.02$ \\
1e & $-0.41\pm0.01$ & $0.31\pm0.01$ & & \\ 
\hline
2a & $-0.24\pm0.01$ & $0.30\pm0.02$ & $-0.21\pm0.01$ & $0.23\pm0.02$ \\
2b & $-0.09\pm0.01$ & $0.23\pm0.02$ & $-0.19\pm0.01$ & $0.18\pm0.01$ \\
2c & $-0.22\pm0.01$ & $0.27\pm0.03$ & $0.04\pm0.01$ & $0.28\pm0.01$ \\
2d & $-0.54\pm0.01$ & $0.24\pm0.01$ & $-0.21\pm0.01$ & $0.10\pm0.02$ \\
2e & $-0.38\pm0.01$ & $0.33\pm0.01$ & & \\ 
\hline
\end{tabular}
\end{center}
\end{table}

As Figure \ref{fig:FitSpect} shows, the agreement between the tilted edge spectrum $F_{\mbox{\scriptsize{tilt}}}$ and the FB-integrated spectrum $F_{\mbox{\scriptsize{int}}}$ is quite good, \ApJMark{especially in the southern bubble,} up to $\sim10\GeV$ energies, beyond which the statistical errors in our spectral measurement become prohibitively large.
The tilt $\myeta$ is consistently stronger in the northern bubble than it is in the southern bubble.
This difference is most extreme at low latitudes (sector d; we cannot measure the tilt in south sector e), yet $F_{\mbox{\scriptsize{tilt}}}$ agrees well with $F_{\mbox{\scriptsize{int}}}$ in both hemispheres even here.

We do not identify significant variations or trends in $\myeta$ among the different sectors within either hemisphere, when using a single edge. 
Averaging $\myeta$ over all available edge 1 sectors thus gives $\myeta=0.27_{-0.02}^{+0.03}$ in the north, and $0.17\pm0.03$ in the south, at $95\%$ confidence levels (CL).
The two hemispheres agree better when using edge 2, where $\myeta=0.28_{-0.03}^{+0.02}$ in the north, and $0.22\pm0.03$ in the south, at $95\%$ CL, possibly due to the shifted high-latitude western sector.
Averaging both edges, we obtain $\myeta=0.28_{-0.02}^{+0.01}$ in the northern bubble and $0.20_{-0.02}^{+0.01}$ in the south, at $95\%$ CL; these are 
shown (as shaded regions) in Figure \ref{fig:DiffusionFit}.

The agreement between $F_{\mbox{\scriptsize{tilt}}}$ and $F_{\mbox{\scriptsize{int}}}$ is naturally explained, in both hadronic and leptonic models, if CRs are injected at the edge and diffuse away. Indeed, as diffusion is expected to become stronger at higher energies, the spectrum of particles remaining in the edge region would thus become softer. To quantify this, we next present a model for CR evolution in the presence of diffusion.

\section{Edge spectrum with diffusion}
\label{sec:Model}

The evolution of CRs in a 'leaky box' model is approximately given by the diffusion-loss equation,
\begin{equation} \label{eq:DiffusionLoss}
\pr_t n = \dot{n}_a + \pr_E\left(\dot{E}_c  n \right) + \grad\cdot (D\grad n) \coma
\end{equation}
where $n(E;\vect{r},t)$ is the differential CR distribution, $\dot{n}_a$ represents the acceleration rate of fresh CRs, $\dot{E}_c$ is the CR power lost to cooling, and $D$ is the CR diffusion function; isotropic diffusion has been assumed.
The second, cooling term on the RHS is important for leptonic models; the analysis remains valid for hadronic models in which it is negligible.
Convective effects, absent in Eq.~(\ref{eq:DiffusionLoss}), are addressed later below.

To obtain a one-zone model, volume integration yields
\begin{equation} \label{eq:DiffusionLossInt}
\pr_t N = \dot{N}_a + \pr_E\left(\dot{E}_c  N \right)   \coma
\end{equation}
where $N\equiv \int n\,dV$ is the volume-integrated distribution.
Here, we neglected the slow volume evolution on the LHS, and discarded surface diffusion terms on the RHS.
This one-zone model can be used to analyze the integrated FB spectrum $F_{\mbox{\scriptsize{int}}}$; see, \eg Gurwich \& Keshet (2017, in preparation). 
\ApJMark{As long as the zone is sufficiently large and diffusion proceeds inward, such that escape from the zone can be neglected,} $F_{\mbox{\scriptsize{int}}}$ \ApJMark{is unaffected by diffusion.}

For steady $\dot{N}_a \propto E^{-p}$ power-law injection and $\dot{E}_c\propto E^2$ (Compton or synchrotron) cooling, this yields the familiar 
\begin{align}
N_0(t) = & \frac{A E^{-p-1}}{(p-1)C}\left[1-\Phi^{p-1}(1-ECt)\right] \\
& \quad \longrightarrow \begin{cases} A t E^{-p} & \mbox{for } E\ll E_c \, ; \\
\frac{A}{(p-1)C}E^{-p-1} & \mbox{for } E>E_c \coma \end{cases}  \nonumber
\end{align}
where $E_c(t)=(Ct)^{-1}$ is the CR cooling break energy, and we took $N(t=0)=0$ as the initial condition.
We also defined the acceleration constant $A\equiv \dot{N}_a E^{p}$, the cooling constant $C\equiv \dot{E}_c E^{-2}$, and the function $\Phi(x)=x\Theta(x)$, with $\Theta(x)$ the Heaviside step function.

Next, consider a zone of finite width near the FB edge, such that diffusion out of the zone does affect the integrated spectrum.
Here, Eq.~(\ref{eq:DiffusionLossInt}) generalizes to
\begin{equation} \label{eq:DiffusionLossIntDiff}
\pr_t N = \dot{N}_a + \pr_E\left(\dot{E}_c  N \right) - \tilde{D}N \coma
\end{equation}
where a tilde represents the loss-surface $\tilde{S}$ at the zone's inner boundary, and we defined the weighted, $\tilde{S}$-averaged diffusion function
\begin{equation} \label{eq:TildeDDef}
\tilde{D}(E)\equiv -\frac{1}{N}\int_{\tilde{S}}D\, d\unit{\sigma}\cdot \grad{n}>0 \fin
\end{equation}

If the CR distribution evolves such that $\tilde{D}$ can be approximated as time-independent, then
\begin{equation}
N \simeq \frac{A E^{-2}}{C}\int_E^{\frac{E}{\Phi(1-E C t)}} e^{-\frac{1}{C}\int_{E}^{E'}\frac{\tilde{D}(E'')}{E''^2}\,dE''}E'^{-p}\,dE' \fin
\end{equation}
This solution shows a spectral break, from an injection--diffusion balanced, $\propto \tilde{D}^{-1}E^{-p}$ spectrum at low energies, to an injection--cooling balanced, $\propto E^{-p-1}$ spectrum at high energies.
In particular, negligible cooling as in the hadronic, or low-energy leptonic, models, yields
\begin{equation} \label{eq:DiffusionWithoutCooling}
N \simeq A \frac{E^{-p}}{\tilde{D}(E)}\left(1-e^{-\tilde{D}t}\right) \coma
\end{equation}
asymptoting to a steady, $N\propto \tilde{D}^{-1}E^{-p}$ distribution at late (as justified below) times.
Diffusion that grows stronger with increasing energy, \ApJMark{as typically expected (and} as assumed henceforth), thus softens the non-cooled spectrum.

For an extended integration zone, CREs necessarily age considerably (by $\tilde{t}$, say) before arriving at $\tilde{S}$.
This may be represented by splitting the zone into a cooling region and a diffusion region, coupled to each other such that the latter is effectively injected with a cooled distribution, $\dot{N}_a\propto N_0(\tilde{t})$. If the diffusive region dominates the signal, then one expects the inferred CR distribution to show a $\propto D^{-1}E^{-p}$ spectrum at low energies, and a $\propto D^{-1}E^{-p-1}$ spectrum at high energies.
It follows that, under such conditions, the integrated CR spectrum is simply softened by $D$ even in the presence of cooling.

The same result is obtained in a model with both convection and diffusion of CRs toward $\tilde{S}$. The averaged CR evolution with increasing, $z>0$ distance from the FB edge \ApJMark{inward} may be modeled by $v\pr_z N=\pr_E(\dot{E}_c N)$, where the mean velocity $v$ includes both convective, $v_c\propto E^0$, and diffusive, $v_d\simeq D(E)/(2z)$, contributions. As $v_d$ declines with \ApJMark{increasing} $z$, it is natural to assume that it dominates near the injection edge, whereas $v_c$ dominates the overall evolution. The result is again an $N\propto D^{-1}E^{-p}$ spectrum at low energies, and a $\propto D^{-1}E^{-p-1}$ spectrum above the cooling break.
Thus, the spectrum resembles the non-diffusive result, uniformly (in $E$) softened by $D$.
Note that here, in the absence of convection, the cooled spectrum would become $\propto E^{-p-1}$, as in the case of a homogeneous zone.

\section{Inferred diffusion function}
\label{sec:Diffusion}

The preceding discussion indicates that if CRs are injected into an edge zone, such as the regions used to compute the spectrum in \S\ref{sec:Spectrum}, but efficiently diffuse away from it, then diffusion can uniformly tilt the edge spectrum with respect to the full integrated FB.

We deduce that \ApJMark{a} typical diffusion \ApJMark{function}, which strengthens as a power law with energy, $D\propto E^\delta$, \ApJMark{may} naturally reconcile $F_{\mbox{\scriptsize{edge}}}$ with the one-zone $F_{\mbox{\scriptsize{int}}}$, as found in Eq.~(\ref{eq:Tilt}), because it can \ApJMark{produce} $F_{\mbox{\scriptsize{edge}}}\propto \epsilon^{-\myeta} F_{\mbox{\scriptsize{int}}}$.
This holds for both hadronic and leptonic models, both with and without cooling.
In a leptonic model, \ApJMark{where an} $N\propto E^{-p}$ \ApJMark{CRE distribution gives rise to an} $F\propto \epsilon^{-(p-1)/2}$ \ApJMark{photon spectrum, diffusion will soften the latter by} $\epsilon^{-\delta/2}$, \ApJMark{so} $\delta=2\myeta$ is needed in order to produce an $\epsilon^{-\myeta}$ tilt in $F$.
In a hadronic model, \ApJMark{where an} $N\propto E^{-p}$ \ApJMark{CRI distribution gives rise to an} $F\propto \epsilon^{-p}$ \ApJMark{photon spectrum, diffusion softens the latter by} $\epsilon^{-\delta}$, and the same \ApJMark{tilt} requires $\delta=\myeta$.

Averaging over all sectors, both hemispheres, and both edges 1 and 2, we find that $\myeta= 0.24\pm0.01$ ($95\%$ CL).
The corresponding energy power-law index $\delta$ of the diffusion function equals $\myeta$ in the hadronic model, whereas in the (more plausible; see \PaperTwo) leptonic model it becomes
\begin{equation}
\delta = 2\myeta = 0.48\pm0.02\, (95\%\mbox{ CL}) \fin
\end{equation}
The full FB-integrated results are similar for the two edges used ($\myeta=0.23_{-0.01}^{+0.02}$ with edge 1, and $\myeta=0.25_{-0.01}^{+0.02}$ with edge 2; $95\%$ CL), as in the hemispheric means of \S\ref{sec:Tilt},
indicating that the measurement is robust.

We do find a significant difference between the values of $\myeta$ inferred in each hemisphere ($\myeta=0.27\pm0.01$ in the north, and $\myeta=0.20\pm0.02$ in the south; $95\%$ CL), and mild variations among the 4--5 sectors within each hemisphere.
These may reflect different evolutionary stages of the magnetic turbulence mediating the CR diffusion.
Another possibility is that the $\myeta$ variations arise from different projection conditions and shock surface morphologies.
Projection effects tend to diminish $\myeta$, by diluting edge regions of strong diffusive CR escape with regions farther from the edge.
Hence, higher values of $\myeta$, as seen in the northern hemisphere, are likely to more accurately represent the diffusion function.

For diffusion to be significant in the edge zone of the FB, it should operate on $\tilde{z}\gtrsim r/10 = 0.5r_5\kpc$ length scales, and over a time scale $t\lesssim 3t_3\Myr$.
Here, $r\equiv 5r_5 \kpc$ is the radius of the FB edge about its approximate center, and we assumed simple ballistic expansion of the FBs over their age $t$.
Hence, we expect $D \gtrsim 3\times 10^{28}r_5^2 t_3^{-1}\cm^2 \se^{-1}$ at $E\sim 10\GeV$ CR energies, which correspond roughly to $\epsilon\sim 1\GeV$ photons in both hadronic and leptonic models, as shown below.

In the leptonic model, the simultaneous manifestation of diffusive and cooling effects has additional implications.
Consider CREs near the cooling-break energy,
\begin{equation}
E_{br} \simeq 10 \left( \frac{\epsilon_{br}/1\GeV}{\epsilon_0/0.4\eV} \right)^{1/2} \GeV \equiv 10 E_{10} \GeV \coma
\end{equation}
where $\epsilon_{br} \simeq 1\GeV$ is the photon cooling break energy, and we assumed Compton cooling off a $U\equiv 1 U_1\eV \cm^{-3}$ starlight photon field of mean photon energy $\epsilon_0\simeq 0.4\eV$.
If the diffusion term in Eq.~(\ref{eq:DiffusionLossIntDiff}) were much stronger than the cooling term at these energies, the cooling break would be shifted to much higher energies, where cooling does become sufficiently strong.
On the other hand, if diffusion were much weaker than cooling near $E_{br}$, the cooled spectrum above the break would be unaffected by diffusion, and show no tilt.

The diffusion and cooling terms are equal at a given energy $E$ if
\begin{equation}
D(E) \simeq \frac{\dot{E}_c r^2}{(\bar{p}-2)E}  \coma
\end{equation}
where $\bar{p}$ is the averaged spectral index of the CREs at $E$.
Imposing this requirement just below the spectral break, where $\bar{p}=p+\delta\simeq 2.1+0.5=2.6$ \citep[using $p=2.1$ inferred from the microwave haze; e.g.][]{PlanckHaze13}, gives $D\simeq 5 \times 10^{29} U_1 r_5^2 E_{10} \cm^2 \se^{-1}$.
Just above the spectral break, where $\bar{p}=p+1+\delta\simeq 3.6$, the same requirement gives $D \simeq 2 \times 10^{29} U_1 r_5^2 E_{10} \cm^2 \se^{-1}$.
We infer that the value of $D(E\simeq E_{br})$ lies near these two estimates.

In the convection--diffusion picture, the diffusive velocity $v_d$ at the inner edge of the zone should be somewhat smaller than the convective velocity. This gives
\begin{equation}
D < 2\tilde{z} v_c \simeq \frac{r^2}{6t} \simeq 4\times 10^{29} r_5^2 t_3^{-1} \cm^2\se^{-1} \coma
\end{equation}
where we assumed that the FB edge is a strong shock with a compression ratio of $4$, as inferred from the microwave spectrum and from the minute variations in $\myeta$, as discussed in \S\ref{sec:Discussion} below.

The above estimate and constraints are consistent with each other within their systematic uncertainties, indicating that our spectral analysis is quite feasible.
The conclusion is that, to within a factor of a few,
\begin{equation} \label{eq:OurD}
D\simeq 3\times 10^{29}(E/10\GeV)^{0.48\pm0.02} \cm^2\se^{-1} \fin
\end{equation}
We may now use this result to confirm that the efficient diffusion condition $\tilde{D}t>1$, used above (\cf Eq.~(\ref{eq:DiffusionWithoutCooling})) to infer the $D^{-1}$ softening \ApJMark{of the CR spectrum,} is satisfied.
For an edge zone of width $\Delta r/r\simeq 0.1$, this requires a mass-density gradient $d \ln \rho/d\ln r>0.8r_5^2t_3^{-1}E_{10}^{-1/2}$ near the edge; see Eq.~(\ref{eq:TildeDDef}).
Therefore, the condition is indeed satisfied, even for edge density gradients as shallow as that in the Primakoff \citep{CourantFriedrichs48, Keller56} solution.

Our results are broadly consistent with typical estimates of the diffusion function and its energy dependence, such as
$D(E)\simeq 5\times 10^{28}E_{10}^{1/2}\cm^2\se^{-1}$ computed for Kraichnan-like turbulence \citep[\eg][]{BerezinskiiEtAl90Book,PtuskinEtAl09};
$D(10\GeV)\simeq 10^{29}\cm^2\se^{-1}$ inferred from the decay of radioactive isotopes \citep{WebberSoutoul98};
and $D(10\GeV)\simeq 6\times 10^{28}\cm^2\se^{-1}$ inferred from GC ridge \gama-rays \citep[][extrapolated from HESS $10^{12.5}\eV$ data using $\delta=0.5$]{DimitrakoudisEtAl09}.
Conversely, these estimates indicate that diffusion and convection should indeed play an important role in shaping the FB edge spectrum.
The \ApJMark{above} estimates slightly differ from each other, and are somewhat lower than estimated in Eq.~(\ref{eq:OurD}); however, \ApJMark{this may be expected, as} diffusion downstream of the FB edges should be modified by the young magnetic fields injected or amplified at the shock.

\section{Summary and Discussion}
\label{sec:Discussion}

We analyze the edges of the FBs using the $\sim 8$ year, Pass 8 \emph{Fermi}-LAT data.
Gradient filters of width $\psi\simeq(2\dgr$--$8\dgr)$ easily pick up the edges (see Figure \ref{fig:Edges}), with little dependence upon $\psi$ in most sectors.
Edge contours identified on two different scales, labeled edge 1 and edge 2, are defined (see Table \ref{tab:EdgeSummary}) and used to study the transition across the FB edge (see Figure \ref{fig:RadialProfiles}).

Both edges connect smoothly, both in location and in orientation, to the intermediate-latitude \emph{ROSAT} X-ray bipolar structure emanating from the GC \citep{BlandHawthornAndCohen03}.
This provides the strongest indication todate that the FBs are a Galactic-scale phenomenon arising from energy release near the GC, and not a small nearby structure seen in projection.

The only significant differences between edges 1 and 2 are found \ApJMark{in} the high-latitude, western part of both north and south bubbles \ApJMark{(in sectors c)}.
These differences arise from nearly linear, somewhat thinner edges picked up in edge 2, diagonally truncating each bubble or part of its flux. The similar, symmetric appearance of these features in the two hemispheres suggests that they are real and significant; they may be related to substructure reported inside the FBs \citep{SuFinkbeiner12}.

Next, we use these edges to perform an assumption-free measurement of the projected FB spectrum in the near-edge region, without using any Galactic model or template.
The results, shown (see Figure \ref{fig:DiffSpect}) in five different sectors (defined in Table \ref{tab:EdgeSectors} \ApJMark{and depicted in Figure} \ref{fig:ThetaBins}) \ApJMark{in each hemisphere}, have far less systematic uncertainties than template-based measurements, albeit larger statistical errors due to the fewer photons collected in the smaller, near-edge regions.
\ApJMark{Thick} (3--4)$\dgr$, \ApJMark{gapped radial bins yield a robust spectral measurement, showing little difference between edges 1 and 2, except in sector c as discussed above. This is confirmed by the convergence and parameter sensitivity tests of Appendix \ref{sec:Convergence}.}

We also examine two edge contours identified manually (edge 3 -- by eye, and edge 4 -- \ApJMark{by} \Su).
These contours are found to be somewhat misplaced and misaligned with respect to the FB edges, suggesting a considerable potential for confusion with substructure, mainly due to \ApJMark{brightness} transitions sharper than the FB edges.
Nevertheless, the manual edges lead to \ApJMark{qualitatively} similar spectral shapes \ApJMark{in most sectors, further} indicating that our measurement of the spectral softening is robust.

The spectra $F_{\mbox{\scriptsize{edge}}}$ we find near the FB edges have a \ApJMark{mildly but} significantly softer slope than the full FB-integrated spectrum $F_{\mbox{\scriptsize{int}}}$.
The two \ApJMark{photon} spectra agree with each other well if a power-law energy tilt (see Eq.~\ref{eq:Tilt}) is introduced, as shown in Figure \ref{fig:FitSpect}.
The averaged slope of the tilt, $\myeta\simeq0.24\pm0.01(95\%\mbox{ CL})$, shows little variations among sectors within each hemisphere, and little dependence upon the choice of edge; see Table \ref{tab:FitParams} and Figure \ref{fig:DiffusionFit}.
It is slightly but significantly larger in the northern hemisphere ($\myeta=0.27\pm0.01$; $95\%$ CL) than in the south ($\myeta=0.20\pm0.02$; $95\%$ CL), suggesting different properties of the underlying magnetic turbulence.

The most natural interpretation of these results is that CRs are injected near the edge, and diffuse away from it \ApJMark{preferentially} at higher energies (see \S\ref{sec:Model}).
The observed softening requires a fairly strong diffusion, $D(E\simeq 10\GeV)\gtrsim 10^{28.5}r_5^2t_3^{-1}\cm^2\se^{-1}$, \ApJMark{consistent with} previous estimates.
A leptonic model \ApJMark{with an} $\epsilon\sim1\GeV$ \ApJMark{cooling break} is more restrictive, because diffusion should soften the spectrum both above and below the break; this gives $D\simeq 10^{29.5}E_{10}^{0.48\pm0.02}\cm^2\se^{-1}$ (see \S\ref{sec:Diffusion}).

Alternative explanations exist, but seem far less plausible.
In leptonic models, the CRE spectrum softens due to cooling, so one may suggest that the CRE population is injected within the bubble, and softens as it migrates out toward the edge.
Alternatively, one may attribute the effect to inverse-Compton emission becoming softer with increasing distance from the GC, due to the increasing extinction of the Galactic radiation field in the background of the fixed CMB.
However, neither scenario should, under reasonable assumptions, produce the observed modest steepening, extending over two orders of magnitude in photon energy, and seen in both intermediate and high latitudes.

The diffusion function we infer from the FB edges \ApJMark{in the leptonic model} is consistent with, and slightly stronger than, previous estimates and indirect measurements (see \S\ref{sec:Diffusion}), as one might expect in the near \ApJMark{downstream of a} shock.
Its energy dependence is consistent with Kraichnan turbulence, but not with Kolmogorov turbulence, often invoked as competing models \citep[\eg][]{NgEtAl10}.
The large extent of the FBs \ApJMark{thus} renders them exceptionally useful for resolving the spacetime evolution of CRs.
Consequently, our inferred $D$, and in particular its energy dependence, are exceptionally well constrained.
We conclude that there is an elevated level of diffusion, with a Kraichnan-like magnetic spectrum, downstream of the shock.

The diffusive softening interpretation is valid only if CRs are injected at the FB edge, unlike some previous suggestions for injection near the GC.
This also indicates that the FB edge is a shock, and not a discontinuity as previously suggested.
Moreover, we find very modest variations in the spectral index correction $\myeta$ among the different sectors, and insignificant variations in $\myeta$ within each hemisphere.
This indicates, for CR injection at the \ApJMark{shock}, that the injection spectrum is nearly constant across the FB edges.
However, one would expect substantial variations in Mach number across the edge, locally due to substructure, but more importantly, globally due to the aspherical bubble morphology. Assuming at least $25\%$ variations in Mach number, the $\Delta p=\pm0.1$ \ApJMark{CRE spectral variations} inferred from $\Delta\myeta=\pm0.05$ imply a minimal Mach number $M\gtrsim 5$, where we assumed Fermi acceleration in a gas of adiabatic index $\gamma=5/3$.
Furthermore, the lack of a clear trend in $\myeta$, except for a slight hint of harder injection at higher latitudes in the northern bubble, suggests that the Mach numbers are even higher than this lower limit.
Note that such a strong shock is inconsistent with the termination shock of a galactic wind, as previously suggested, indicating that the FB edges are in fact a forward shock.

These results are consistent with the strong shock inferred from the microwave haze. The low Mach numbers derived from the \ion{O}{8}/\ion{O}{7} line strength ratio or from the high-latitude $\sim0.3\keV$ X-ray component are \ApJMark{then} likely due to a combination of foreground confusion, projection effects, and dilution by surrounding regions.

\acknowledgements
We thank D. Finkbeiner for encouragement and helpful advice.
N. Sherf, T. Gur, I. Reiss, and R. Crocker are acknowledged for assistance and interesting discussions.
This research (grant No. 504/14) was supported by the ISF within the ISF-UGC joint research program, received funding from IAEC-UPBC joint research foundation grant 257, and was supported by the GIF grant I-1362-303.7/2016.

\bibliography{FermiBubbles}

\appendix

\section{Convergence tests}
\label{sec:Convergence}

\subsection{Dependence on the gap separation $\theta_g$ and edge location $\theta_0$}

\begin{figure*}
\PlotFigs{
\centerline{
\begin{overpic}[width=\figsizeB]{\myfig{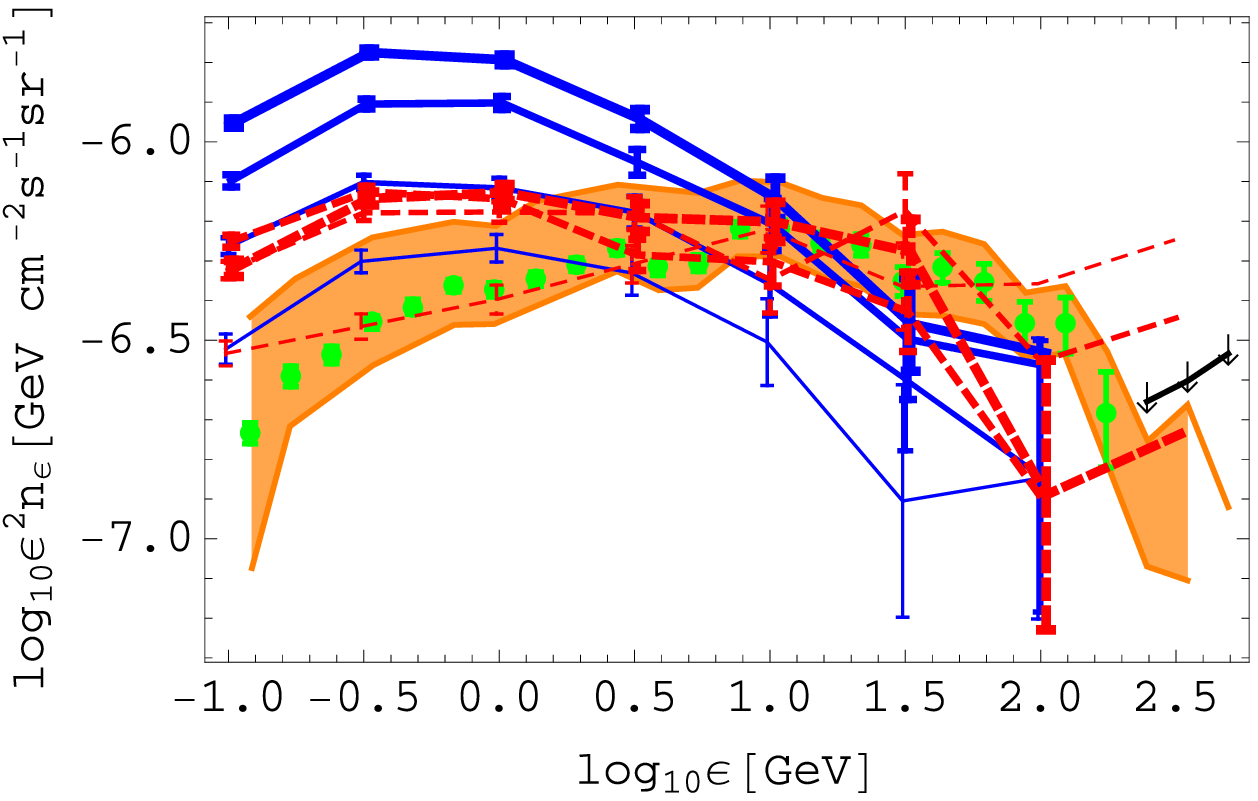}}
 \put (90,57) {\normalsize \textcolor{black}{a}}
\end{overpic}
}
\vspace{-2.7cm}
\centerline{
\begin{overpic}[width=\figsizeB]{\myfig{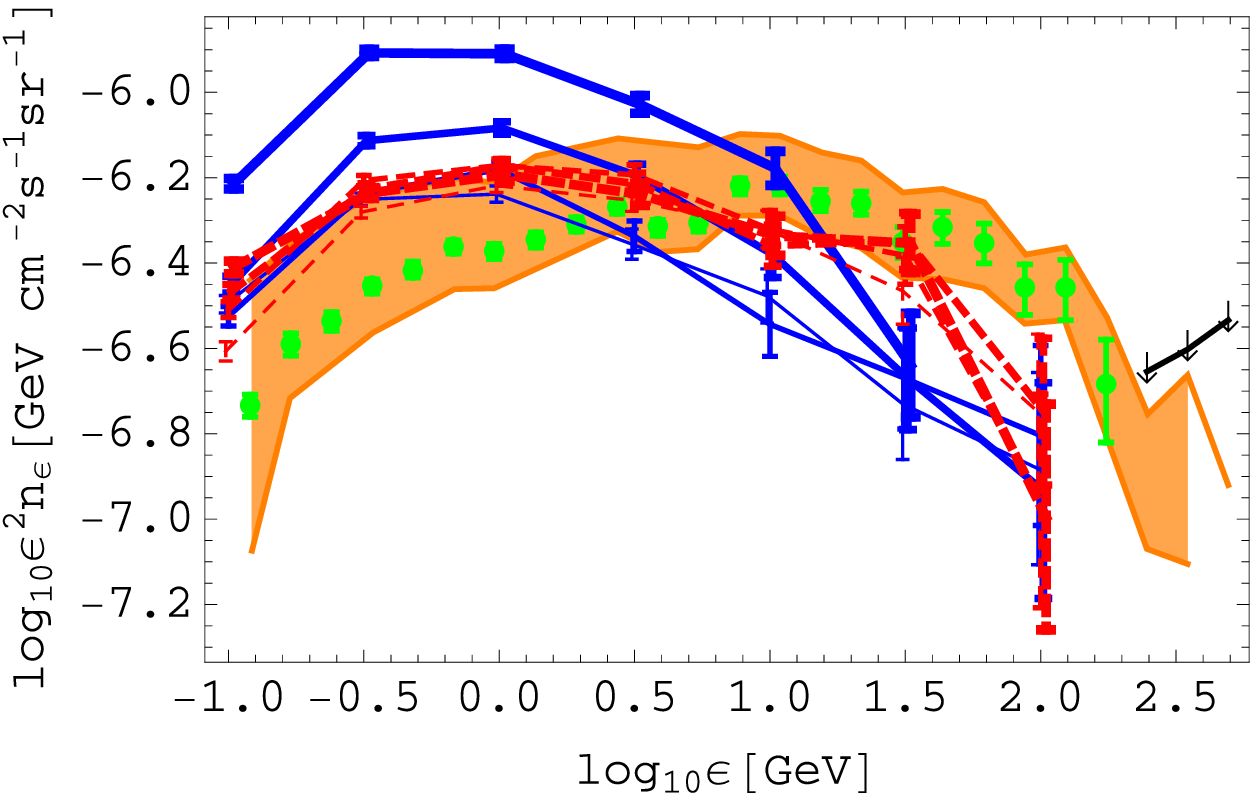}}
 \put (90,57) {\normalsize \textcolor{black}{b}}
\end{overpic}
\hspace{\figsizeB}
\begin{overpic}[width=\figsizeB]{\myfig{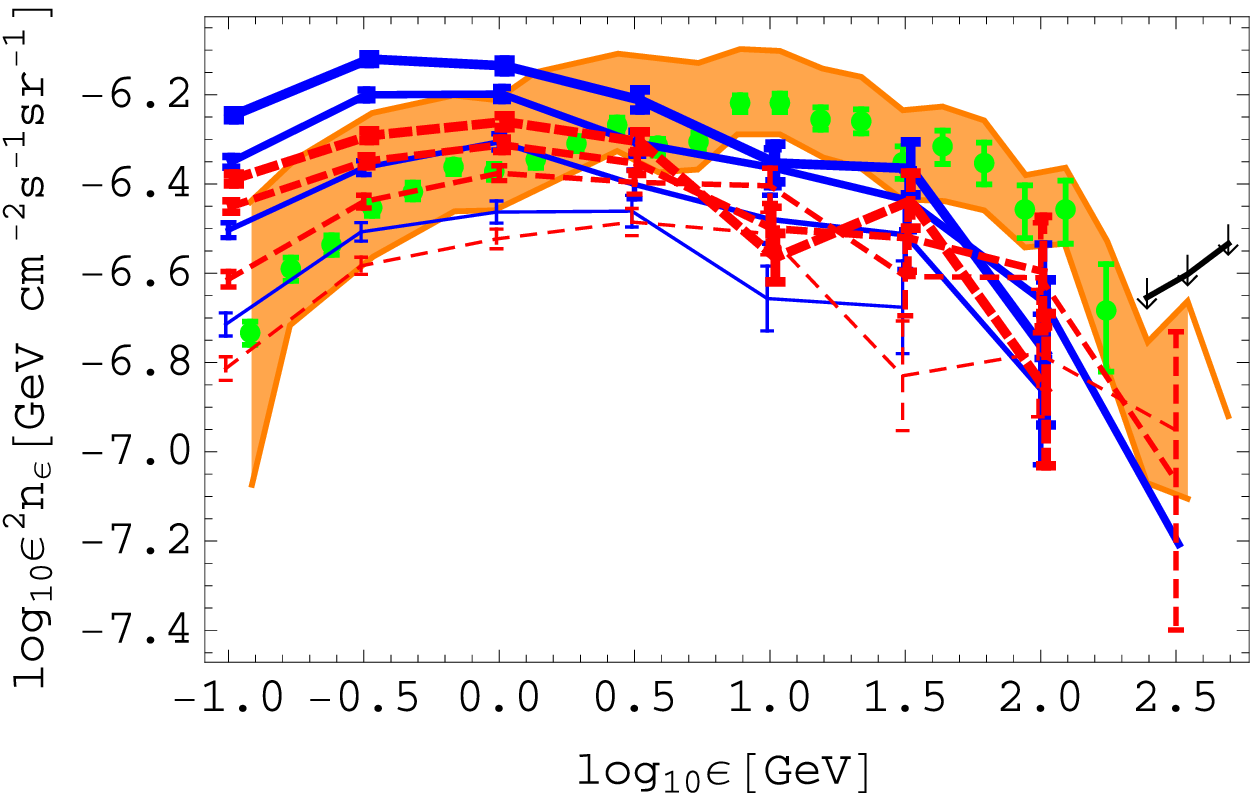}}
 \put (90,57) {\normalsize \textcolor{black}{c}}
\end{overpic}
}
\vspace{0.3cm}
\centerline{
\begin{overpic}[width=\figsizeB]{\myfig{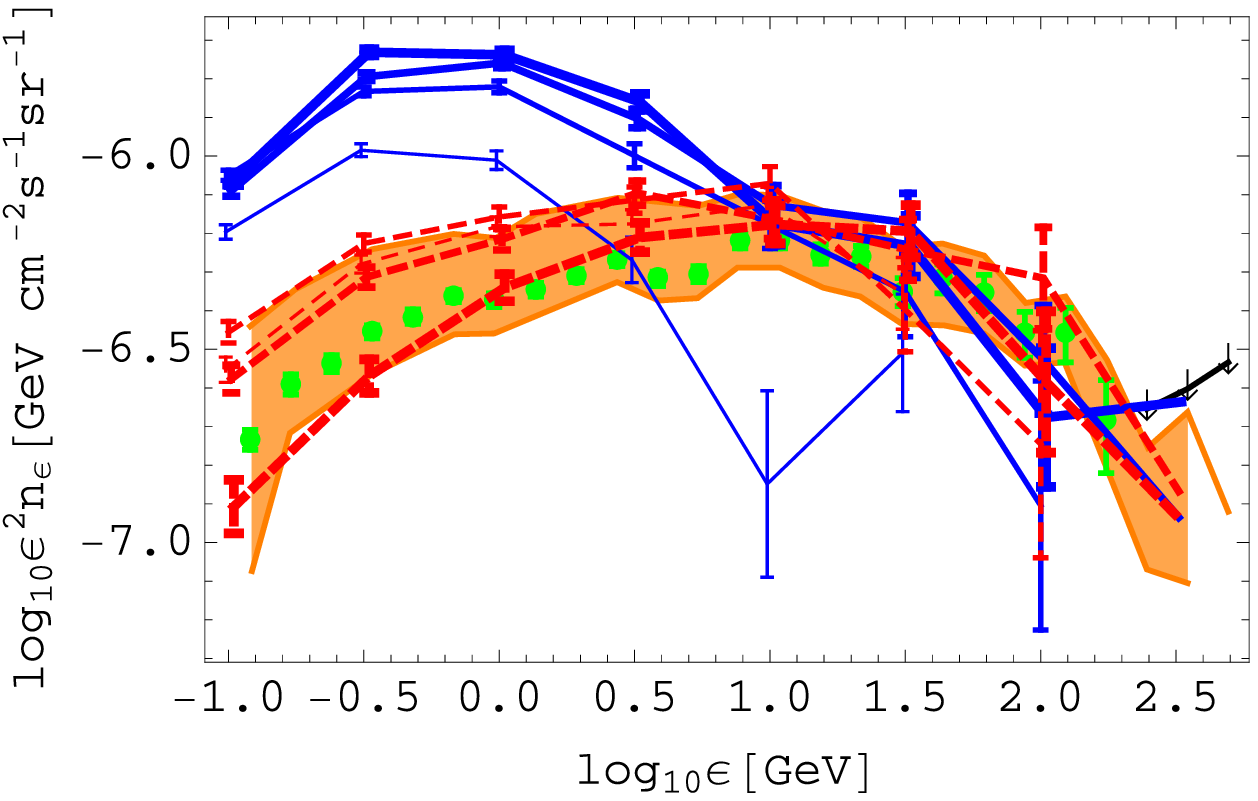}}
 \put (90,57) {\normalsize \textcolor{black}{d}}
\end{overpic}
\begin{overpic}[width=\figsizeB]{\myfig{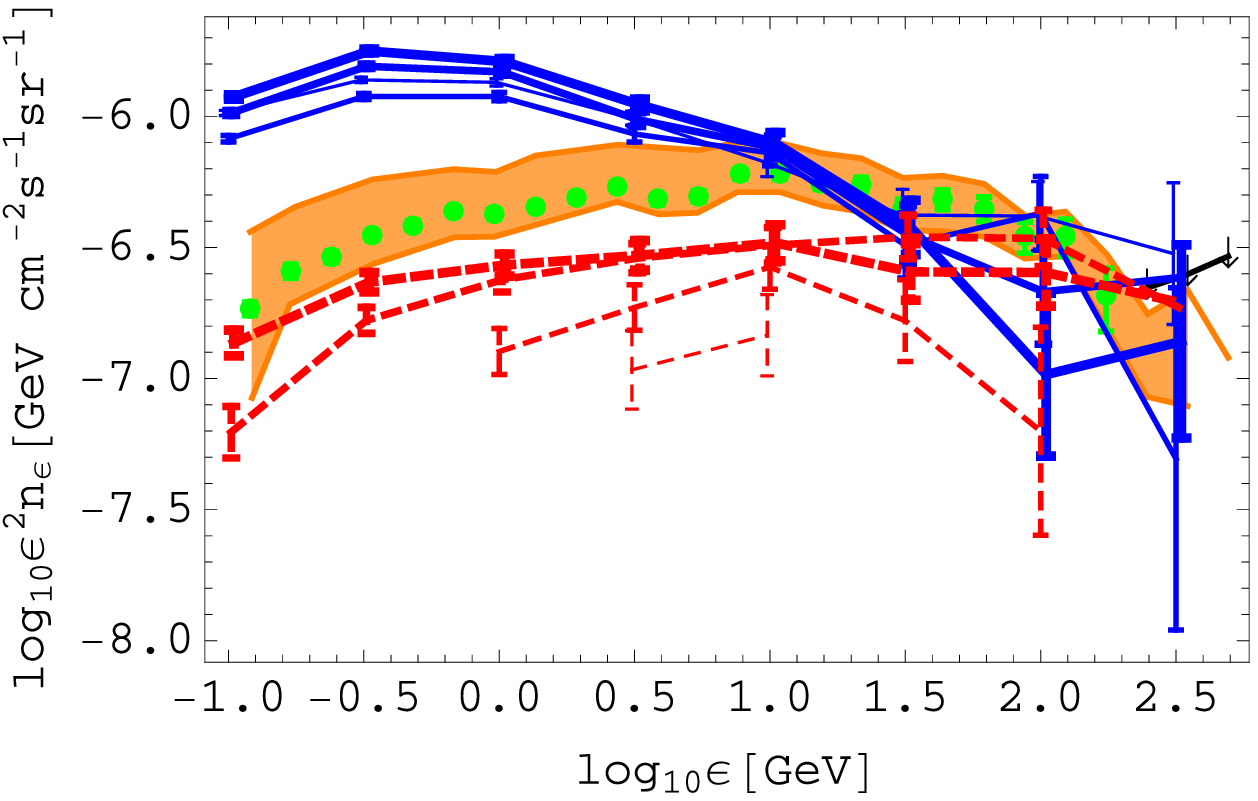}}
 \put (90,57) {\normalsize \textcolor{black}{e}}
\end{overpic}
}
}
\caption{ \label{fig:CGvsGap}
Same as Figure \ref{fig:DiffSpect}, but focusing on edge 1 only, and varying the gap size as $\theta_g=3\dgr,6\dgr,9\dgr,12\dgr$ (thin to thick curves).
}
\end{figure*}

\begin{figure*}
\PlotFigs{
\centerline{
\begin{overpic}[width=\figsizeB]{\myfig{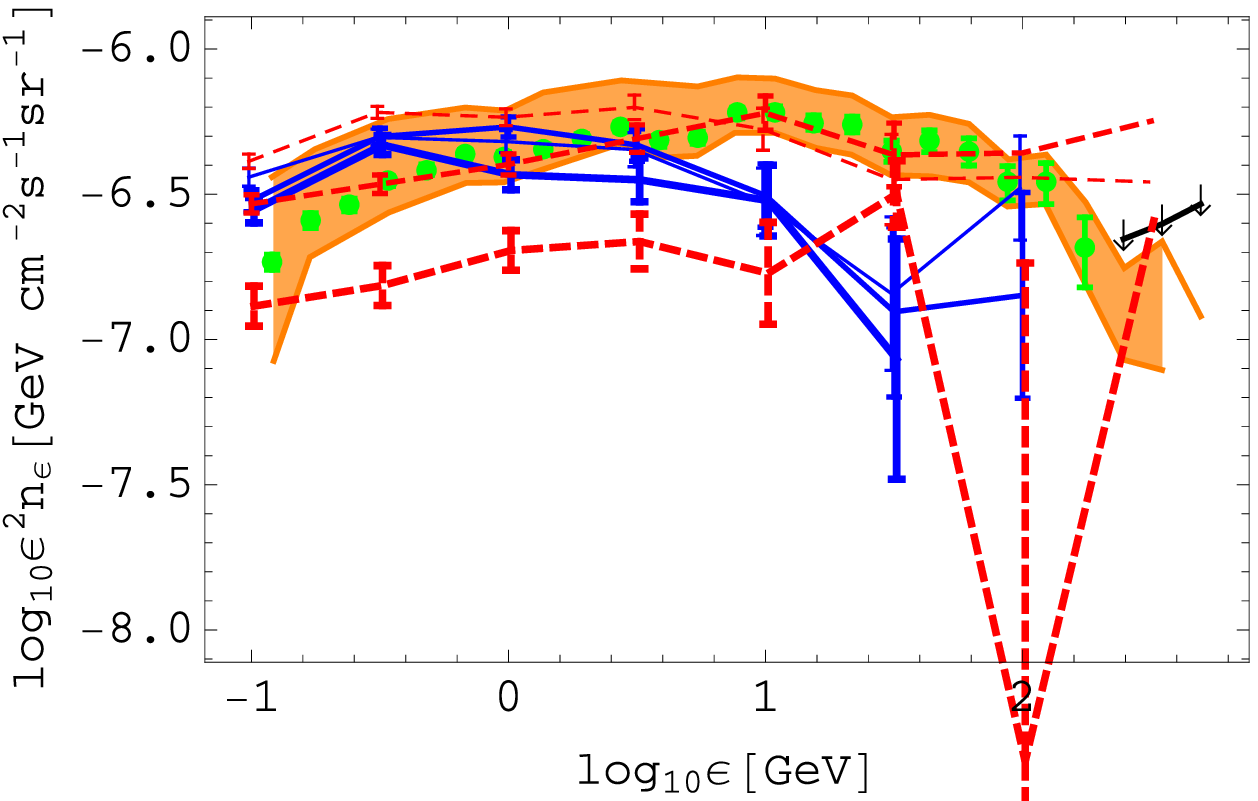}}
 \put (90,57) {\normalsize \textcolor{black}{a}}
\end{overpic}
}
\vspace{-2.7cm}
\centerline{
\begin{overpic}[width=\figsizeB]{\myfig{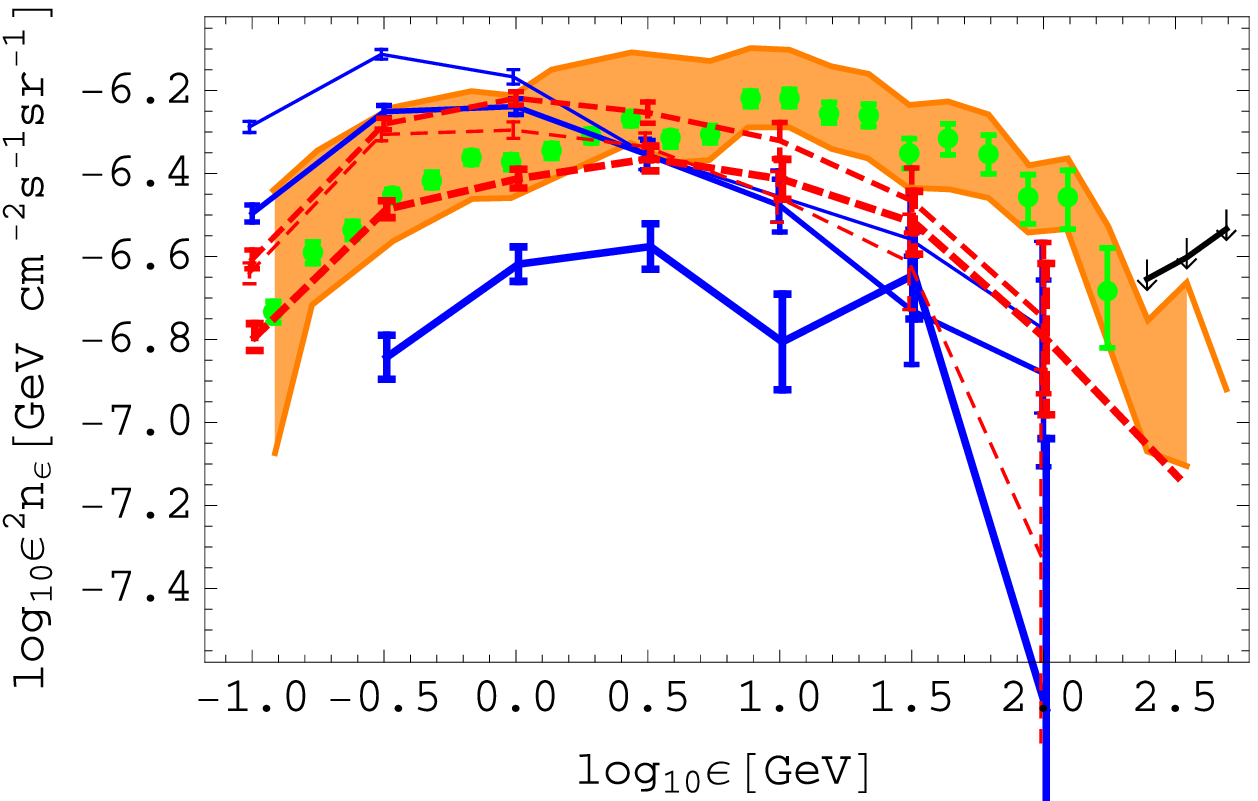}}
 \put (90,57) {\normalsize \textcolor{black}{b}}
\end{overpic}
\hspace{\figsizeB}
\begin{overpic}[width=\figsizeB]{\myfig{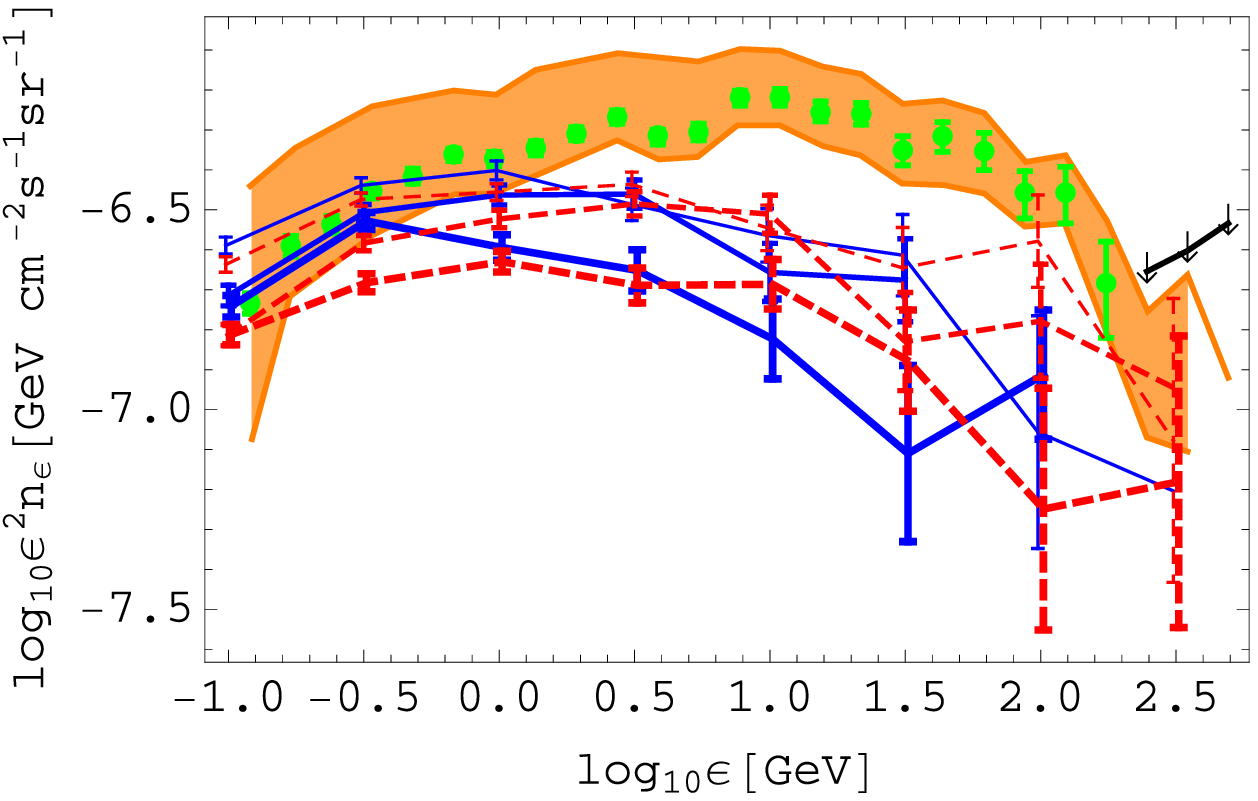}}
 \put (90,57) {\normalsize \textcolor{black}{c}}
\end{overpic}
}
\vspace{0.3cm}
\centerline{
\begin{overpic}[width=\figsizeB]{\myfig{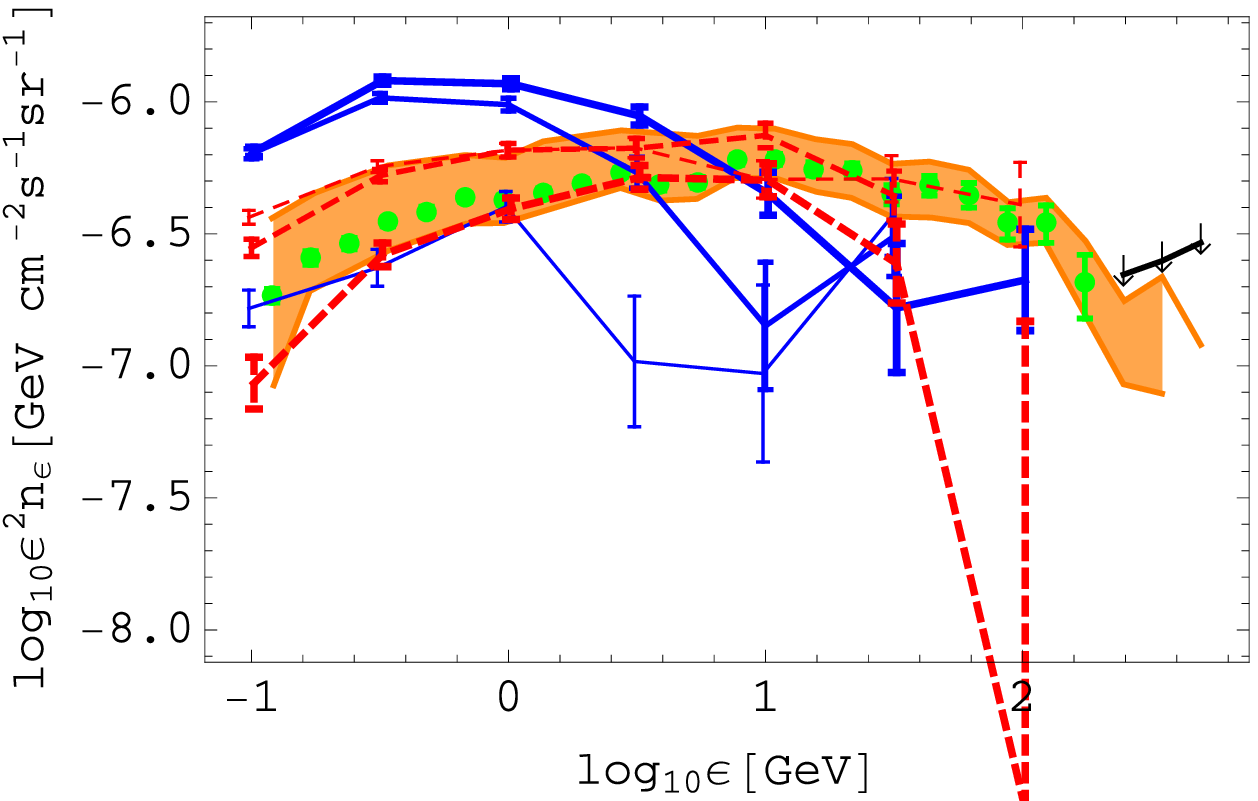}}
 \put (90,57) {\normalsize \textcolor{black}{d}}
\end{overpic}
\begin{overpic}[width=\figsizeB]{\myfig{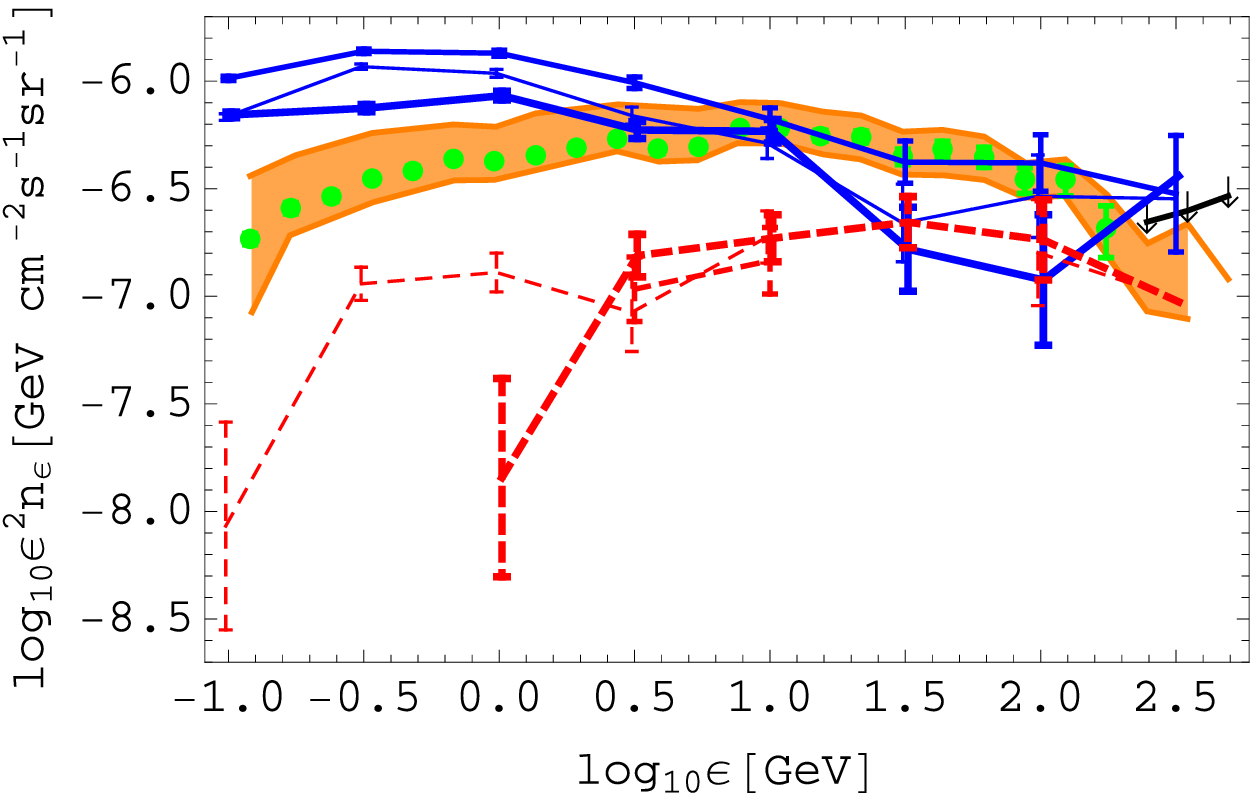}}
 \put (90,57) {\normalsize \textcolor{black}{e}}
\end{overpic}
}
}
\caption{ \label{fig:CGvsOffset}
Same as Figure \ref{fig:DiffSpect}, but focusing on edge 1 only, using $\theta_g=3\dgr$, and offsetting $\Delta\theta$ by $\theta_0=-2\dgr,0\dgr,+2\dgr$ (thin to thick curves).
}
\end{figure*}

The method we used to derive the spectrum depends on the precise location of the edge, denoted as $\theta_0=0$, the angular gap $\theta_g$ between the inner and outer regions, and the width $\theta_r$ of each region.
\ApJMark{These scales are chosen to be much larger than the \emph{Fermi}-LAT point spread function (PSF) at the relevant energies, in order to prevent the smearing of the signal; note that the spectral softening we find is opposite to the putative effect of such contamination. }
Here, we test the dependence of the results on these parameters. For \ApJMark{simplicity}, we focus on edge 1.
Figure \ref{fig:CGvsGap} shows the spectrum inferred for different choices of $\theta_g$, and Figure \ref{fig:CGvsOffset} shows the dependence of the spectrum on small changes in $\theta_0$.

As Figure \ref{fig:CGvsGap} shows, in the southern hemisphere the results are converged by $\theta_g=6\dgr$ for the east sectors (a, b, and d), whereas sector c suggests that the edge is smeared, probably due to a misaligned projection in the southwest.
In the northern hemisphere, the low-latitude sectors (d and e) are well converged by $\theta_g=6\dgr$, but the high-latitude sectors (a, b, and c) show evidence for a smeared edge, in particular in the west.

Figure \ref{fig:CGvsOffset} indicates that the spectrum is not sensitive to changes of $\lesssim2\dgr$ in $\theta_0$. For larger misalignments, the estimated edge flux declines, indicating that the \ApJMark{initial} edge positioning is fairly good.

Notice that even in the poorly projected sectors, c in the south and a--c in the north, the spectral shape shows little variations with $\theta_g\gtrsim6\dgr$ and with $|\theta_0|\lesssim 2\dgr$. We conclude that our analysis and spectral conclusions are robust to small changes in these parameters, although the normalization factors $A$ (see Table \ref{tab:FitParams}) do somewhat depend on the method parameters \ApJMark{used} in these poor-projection sectors.

Figure \ref{fig:CGvsGap} shows a similar dependence upon $\theta_g$ in different energy bands. We therefore find no evidence that the FBs are more extended at high energies, as reported by \cite{YangEtAl14}.

\subsection{Extrapolation to the edge}

Our method of computing the spectrum by taking the difference between regions inside and outside the FB edge, $F_{\mbox{\scriptsize{edge}}}=F(S_i)-F(S_o)$, is susceptible to some degree of foreground contamination, as the regions are extended, with angular width $\theta_r$, and separated by the gap $\theta_g$, \ApJMark{whereas the foreground may vary across such scales.}
To test for such contamination, we model the dependence of each region as a power law in $\Delta\theta$, extrapolate both results to the edge, and compute their difference.

If foreground contamination is substantial, this should yield a spectrum very different from that of the direct differentiation method of Figure \ref{fig:DiffSpect}.
For foreground and FB emission that smoothly vary with $\Delta\theta$, this method would remove most of the putative foreground contamination in our results.
In contrast, ill behaved signals may produce a substantial scatter in the results.

\begin{figure*}
\PlotFigs{
\centerline{
\begin{overpic}[width=\figsizeB]{\myfig{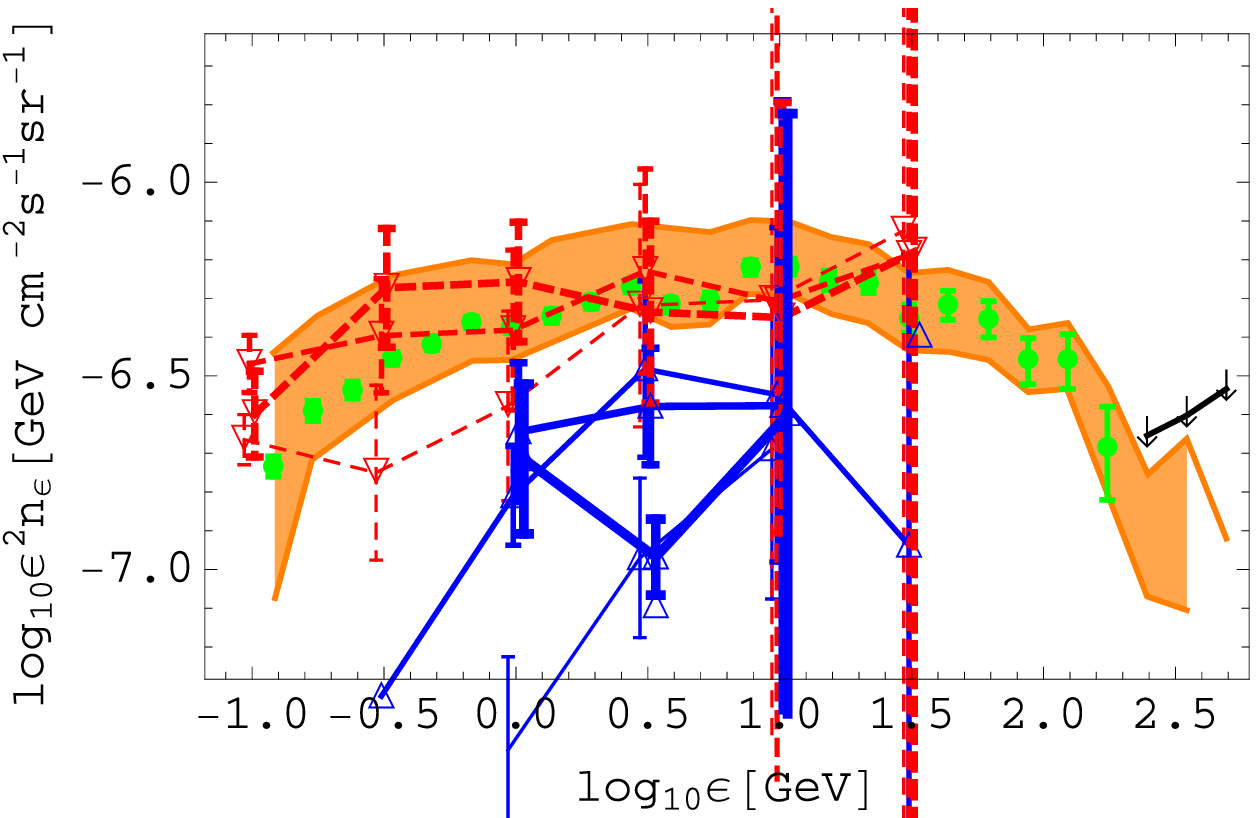}}
 \put (90,57) {\normalsize \textcolor{black}{a}}
\end{overpic}
}
\vspace{-2.7cm}
\centerline{
\begin{overpic}[width=\figsizeB]{\myfig{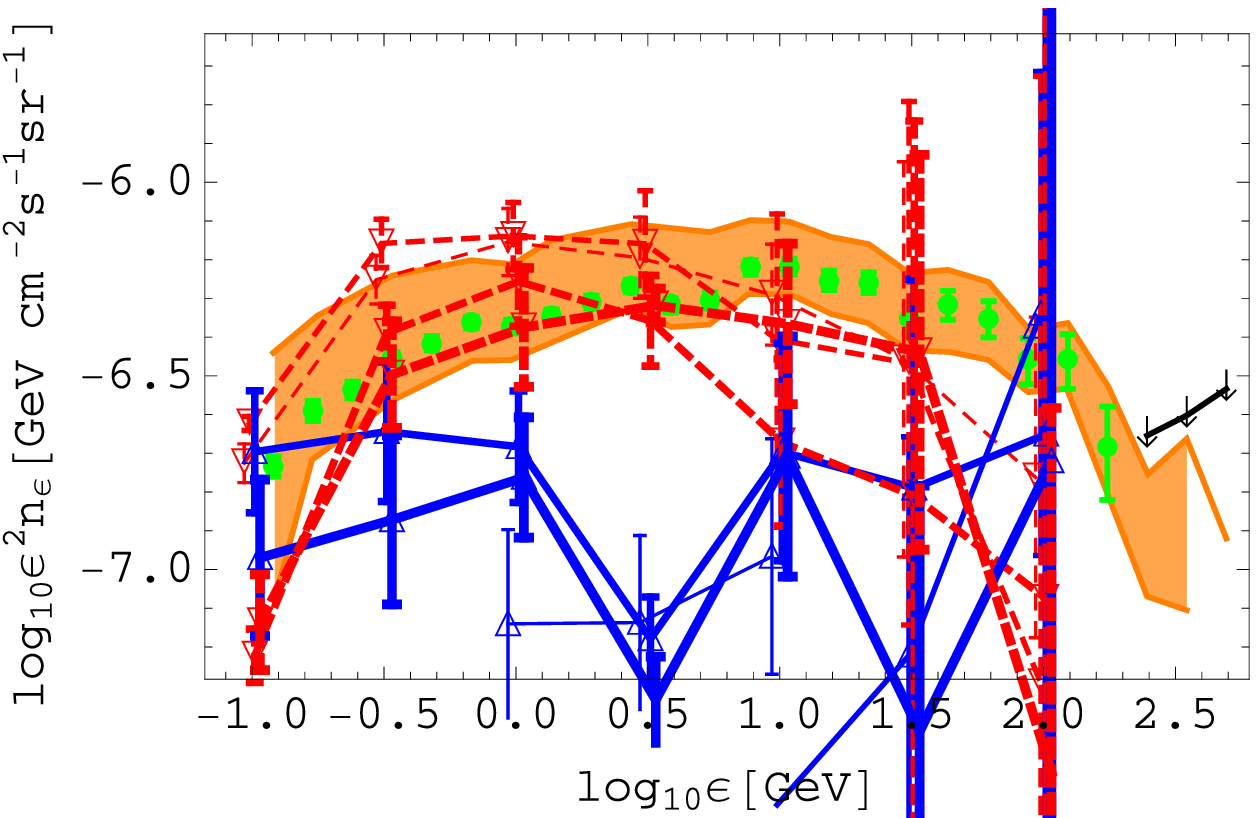}}
 \put (90,57) {\normalsize \textcolor{black}{b}}
\end{overpic}
\hspace{\figsizeB}
\begin{overpic}[width=\figsizeB]{\myfig{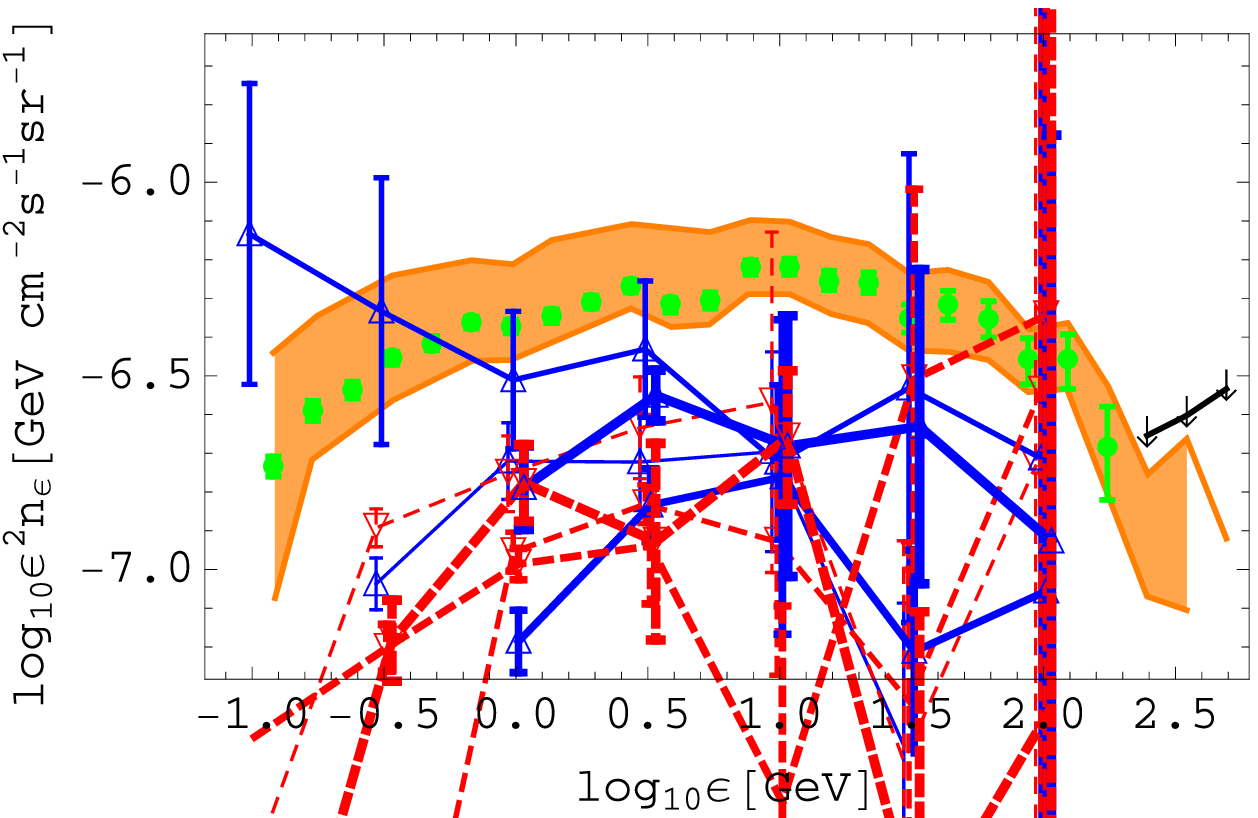}}
 \put (90,57) {\normalsize \textcolor{black}{c}}
\end{overpic}
}
\vspace{0.3cm}
\centerline{
\begin{overpic}[width=\figsizeB]{\myfig{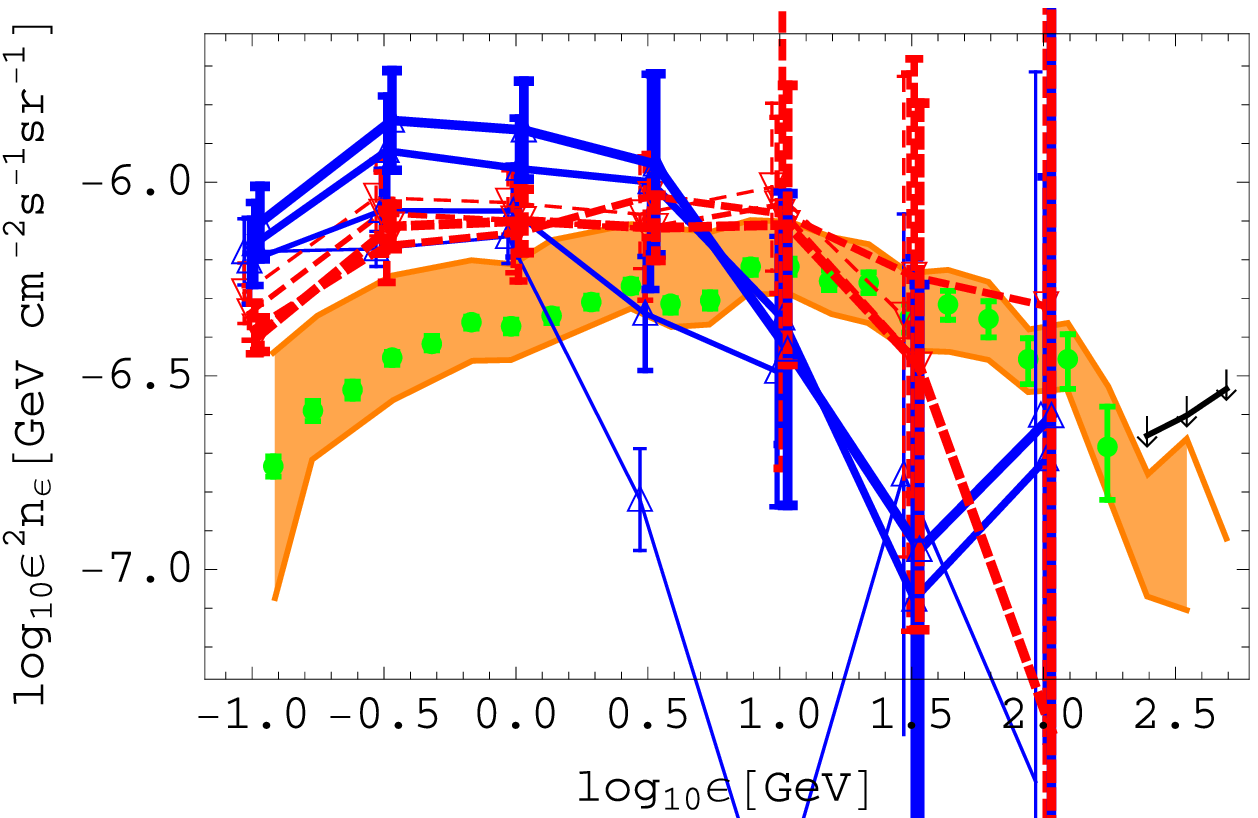}}
 \put (90,57) {\normalsize \textcolor{black}{d}}
\end{overpic}
\begin{overpic}[width=\figsizeB]{\myfig{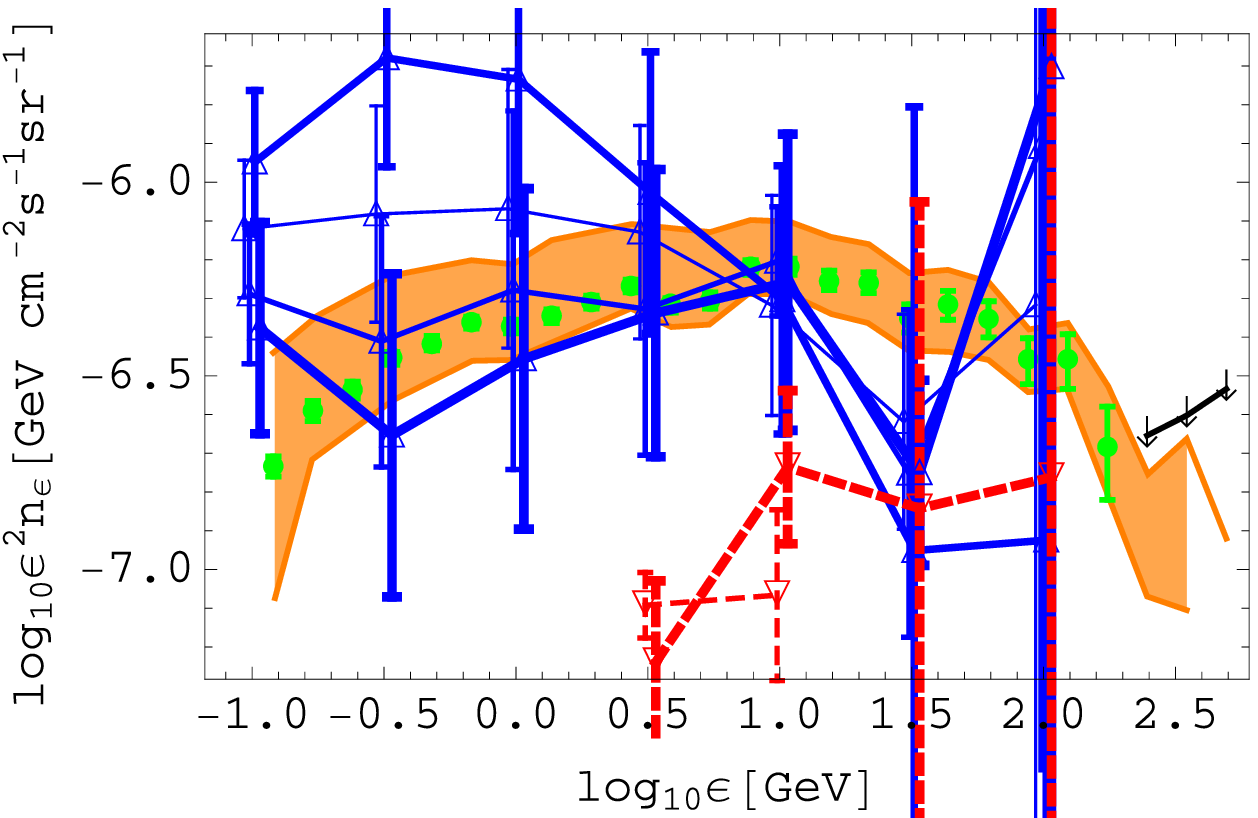}}
 \put (90,57) {\normalsize \textcolor{black}{e}}
\end{overpic}
}
}
\caption{ \label{fig:InterpolatedSpect}
Same as Figure \ref{fig:DiffSpect}, but using the extrapolation method, and $\theta_g=4\dgr$.
}
\end{figure*}

Figure \ref{fig:InterpolatedSpect} shows the results of this extrapolation method.
In the southern hemisphere, which has less interfering structure, the results for sectors a, b, and d are similar to those of the direct difference method, although the statistical uncertainties in the extrapolation are considerable, and at high energies \ApJMark{become} excessive.
In the northern hemisphere, we obtain spectra only at low latitudes (sectors d and e); these results are also qualitatively similar to those of the direct method.
We conclude that foreground contamination in the direct method is unlikely to be severe.


\end{document}